\documentclass[amsmath,amssymb,eqsecnum,notitlepage]{revtex4-1}
\pdfoutput=1
\usepackage{graphicx}
\usepackage{subfigure}
\usepackage[breaklinks, colorlinks, citecolor=blue]{hyperref}
\usepackage{amsmath}
\usepackage{amssymb}
\usepackage{bm}

\everymath{\displaystyle}

\graphicspath{{./}}

\def\half{\frac{1}{2}}
\def\third{\frac{1}{3}}
\def\quarter{\frac{1}{4}}
\def\0{\varnothing}
\def\p{\parallel}
\def\x{\mathsf{x}}

\def\l{\lambda}

\def\Rfour{{}^{(4)}\!R}
\def\Rthree{{}^{(3)}\!R}
\newcommand{\partdif}[2]{\frac{\partial {#1}}{\partial {#2}}}
\newcommand{\funcdif}[2]{\frac{\delta {#1}}{\delta {#2}}}

\newcommand{\figref}[1]{Fig.~\ref{#1}}

\newcommand{\secref}[1]{section~\ref{#1}}
\newcommand{\subsecref}[1]{subsection~\ref{#1}}
\newcommand{\sgn}[1]{\operatorname{sgn} (#1)}
\newcommand{\tr}[1]{\operatorname{Tr} (#1)}

\begin{document}

\begin{flushleft}
KCL-PH-TH/2018-26
\end{flushleft}

\title{Deformed general relativity and scalar-tensor models}

\author{Rhiannon Cuttell\href{https://orcid.org/0000-0002-3327-5313}{\includegraphics[height=0.666\baselineskip]{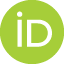}}} \email{rhiannon.cuttell@kcl.ac.uk}
\author{Mairi Sakellariadou} \email{mairi.sakellariadou@kcl.ac.uk}
\affiliation{
    Department of Physics,
	King's College London,
	University of London,
	Strand,
	London,
	WC2R 2LS,
	U.K.
}
\date{\today}

\begin{abstract}
    We calculate the most general action for a scalar-tensor model up to quadratic order in derivatives with deformed general covariance and non-minimal coupling.  We demonstrate how different choices of the free functions recover specific well known scalar-tensor models.  We look at the cosmological dynamics and find the general conditions for either inflation or a big bounce.  Using this we present a novel non-minimally coupled scalar model which produces a bounce, and describe how to find similar models.
\end{abstract}

\maketitle


\tableofcontents


\section{Introduction}
\label{sec:intro}

One of the leading candidates for a theory of quantum gravity is loop quantum gravity, which describes space-time as a network of loops and connections \cite{Rovelli:2014ssa}.  It focuses on maintaining some key concepts from general relativity such as background independence and local dynamics throughout the process of combining gravity and quantum mechanics.
A related theory is loop quantum cosmology, which uses concepts and techniques from the former and applies them directly at the cosmological level by using midi-superspace models \cite{bojowald_quantum_2012, Ashtekar:2011ni}.  That is, by quantising a universe which already has certain symmetries assumed such as isotropy to simplify the process.  There has been some progress towards proving that loop quantum gravity can be symmetry-reduced to loop quantum cosmology but as yet this has not been shown definitively \cite{Gielen:2013kla, Alesci:2016gub}.

Space and time are in principle treated on an equal footing by general relativity, but we must split them and choose a specific time coordinate when seeking to make contact with quantum mechanics or to numerically simulate them.  Even though we are left with an image of a spatial hypersurface evolving through time, the choice of time is kept arbitrary through symmetries embodied in the hypersurface deformation algebra which is of the form $\left\{C_i,C_j\right\}=f_{ijk}C_k$ \cite{dirac1964, bojowald2010canonical}.  This is a Lie algebroid which describes the relationships between the invariant constraints of the spatial manifold in order to maintain the space-time geometry, and generates transformations between different choices of coordinates \cite{hojman_geometrodynamics_1976, Bojowald:2016hgh}.

For models of loop quantum cosmology to be self-consistent and be free of anomalies in the algebra while including some of the interesting effects which come from the discrete geometry, it seems that the algebra of constraints must be deformed so that some of the structure functions become more dependent on the phase space variables through a deformation function $f_{ijk}(q)\to{}\beta(q,p)f_{ijk}(q)$ \cite{Cailleteau2012a, Mielczarek:2012pf}.  Deforming rather than breaking the algebra in principle maintains general covariance but the transformations between different choices of coordinates become highly non-linear and interpretation in terms of classical notions of geometry becomes less clear.
Taking the deformed constraint algebra to the flat-space limit gives us a deformed version of the Poincar\'{e} algebra, leading to a modified dispersion relation \cite{Bojowald:2012ux, Brahma:2016tsq}.  This might be revealing a non-commutative character to geometry \cite{Amelino-Camelia:2016gfx, Bojowald:2017kef} although probably not a multifractional one \cite{Calcagni:2016ivi}.  The deformation function may change sign, making the hyperbolic equations become elliptical, implying a transition from classical Lorentzian space-time to an effectively Euclidean quantum regime \cite{Bojowald:2016hgh, Bojowald:2016vlj} and a running of the spectral dimension \cite{Mielczarek:2016zfz}.

However, there is ambiguity in the correct choice of variables used for loop quantum gravity.  The results cited above are for real variables for which there might be some difficulty including matter and local degrees of freedom \cite{Bojowald:2016itl}.  The self-dual variables have had some positive results for including those degrees of freedom without deforming the constraint algebra \cite{BenAchour:2016brs}, but might not have the desirable quality of resolving curvature singularities \cite{Brahma:2016tsq}.

The semi-classical model called deformed general relativity \cite{bojowald_deformed_2012} was created by taking the deformed constraint algebra and finding a corresponding action using metric variables and including local degrees of freedom \textit{a priori}.  We can do this because if we start from an algebra of constraints and make some assumptions, we can deduce the general form of all the constraints \cite{kuchar_geometrodynamics_1974, hojman_geometrodynamics_1976}.  This should provide a more intuitive understanding of how the deformation affects dynamics and may provide a guide for how to include the problematic degrees of freedom when working with real variables in loop quantum gravity.

In this paper we build on preceding work by ourselves and others \cite{kuchar_geometrodynamics_1974, hojman_geometrodynamics_1976, bojowald_deformed_2012, Cuttell2014} by including a scalar field with non-minimal coupling into this calculation of the effective action up to quadratic order in derivatives.  There is particular emphasis on scalar fields which are of geometric origin such as with parameterisations of $\textstyle{}F\left({}^{(4)}\!R\right)$ gravity \cite{Deruelle:2009pu, Deruelle2010}. We aim to lay the groundwork for a future study where we will calculate a deformed effective action with higher powers of derivatives which cannot be parameterised into a scalar field.

\subsection*{Motivating the ansatz}
\label{sec:def-st_ansatz}

The calculation is simpler when we can provide a well motivated ansatz for the action or Hamiltonian.  So we give here an example of the form of a Hamiltonian constraint we wish to replicate in a more general way before we account for the deformed constraint algebra.  We start from the action,
\begin{equation}
    S := \int \mathrm{d}t \mathrm{d}^3 x L = \half \int \mathrm{d}t \mathrm{d}^3x \sqrt{\left|g\right|} F \left( \Rfour \right),
        \label{eq:action_pure}
\end{equation}
where $L$ is the Lagrangian, $g$ is the determinant of the space-time metric tensor and $\Rfour$ is the four dimensional Ricci curvature scalar.  We are working with the units $G=1/8\pi$ and $c=1$ so that Einstein's gravitational constant is equal to unity.

We wish to use the Hamiltonian formulation because it is necessary for a discussion about the constraint algebra, but it cannot easily deal with time derivatives of second order or higher.  For the Einstein-Hilbert action they are removed by simply integrating by parts, but for more general actions we must introduce extra fields.

The conventional way of dealing with actions containing higher order derivatives is by using the Ostragradsky method, whereby the velocity of a field is itself treated as an independent field.  However, this method can introduce an instability and so is undesirable \cite{Woodard:2006nt}.
Another way of dealing with the higher order time derivatives is by introducing a field which absorbs the higher order dynamics, as shown in \cite{Deruelle:2009pu, Deruelle2010}.  We will outline this process now.  Firstly, we introduce an auxiliary variable $\rho$ to simplify dealing with the higher orders,
\begin{equation}
    S = \half \int \mathrm{d}t \mathrm{d}^3x \sqrt{\left|g\right|} \left\{ F \left( \rho \right) + \psi \left( \Rfour - \rho \right) \right\},
        \label{eq:action_rho}
\end{equation}
where $\psi$ acts as a Lagrange multiplier, ensuring $\rho\approx\Rfour$.  The approximation sign $\approx$ is used to indicate an equality which is true `on-shell', or at the dynamical level.  To proceed, we implement a conventional space-time decomposition \cite{bojowald2010canonical},
\begin{equation}
\begin{gathered}
    q_{\mu\nu} = g_{\mu\nu} + n_\mu n_\nu, 
        \quad
    v_{ab} := \mathcal{L}_n q_{ab},
        \quad
    R = \Rthree,
        \\
    \Rfour  = R + q^{ab} \mathcal{L}_n v_{ab} + \quarter v^2 - \frac{3}{4} \tr{v^2} - \frac{2}{N} \Delta N,
\end{gathered}%
\end{equation}
where $q_{ab}$ is the induced metric field on the spatial hypersurface, and $n^\mu$ is the time-like normal vector.  We use Lie derivatives with respect to the normal vector as velocities because the time vector can be split into its normal and tangential components $t^\mu=Nn^\mu+N^\mu$, where $N$ is the lapse and $N^a$ is the shift, and therefore $\mathcal{L}_nX=N^{-1}\left(\dot{X}-\mathcal{L}_NX\right)$.

From the action \eqref{eq:action_rho} we can integrate by parts to move the second time derivatives to $\psi$ which promotes it to a dynamical variable,
\begin{equation}
    S = \half \int \mathrm{d}t \mathrm{d}^3x N \sqrt{q} \left\{ F \left( \rho \right) + \psi \left[ R - \mathcal{K} - \frac{2}{N} \Delta N - \rho \right] - \nu v \right\},
\end{equation}
where $q:=\det{q_{ab}}$, $\nu:=\mathcal{L}_n\psi$, and $\mathcal{K}:=\left[v^2-\tr{v^2}\right]/4$ is the standard gravitational kinetic term.  The conjugate momenta are,
\begin{subequations}
\begin{align}
    p^{ab}(x) & := \int \mathrm{d}^3 y \funcdif{L(y)}{\dot{q}_{ab}(x)} = \frac{1}{N(x)} \int \mathrm{d}^3 y \funcdif{L(y)}{v_{ab}(x)} = \half \sqrt{q} \left\{ \frac{\psi}{2} v_{cd} \left( Q^{abcd} - q^{ab} q^{cd} \right) - \nu q^{ab} \right\},
        \\
    \Pi (x) & := \int \mathrm{d}^3 y \funcdif{L(y)}{\dot{\psi}(x)} = \frac{1}{N(x)} \int \mathrm{d}^3 y \funcdif{L(y)}{\nu(x)} = \frac{-1}{2} \sqrt{q} v,
\end{align}
        \label{eq:geo_momenta}%
\end{subequations}
where $Q^{abcd}:=q^{a(c}q^{d)b}$ as in \eqref{eq:metric_combinations}.  We can invert this to find,
\begin{equation}
    v_{ab} = \frac{4}{\psi\sqrt{q}} p^{cd} \left( Q_{abcd} - \third q_{ab} q_{cd} \right) - \frac{2}{3\sqrt{q}} q_{ab} \Pi,
        \quad
    \nu = \frac{2}{3\sqrt{q}} \left( \psi \Pi - p \right).
        \label{eq:geo_velocities}
\end{equation}
We Legendre transform the Lagrangian to find the associated Hamiltonian,
\begin{equation}
    H = \int \mathrm{d}^3 x \left( \dot{q}_{ab} p^{ab} + \dot{\psi} \Pi + \mu_\rho \pi_\rho + \mu_N \pi_N + \mu^N_a \pi_N^a - L \right),
\end{equation}
with the corresponding Hamiltonian constraint,
\begin{equation}
    C := \funcdif{H}{N}
    = \frac{2}{\sqrt{q}} \left( \frac{1}{\psi} \tr{p}^2 - \frac{1}{3\psi} p^2 - \third p \Pi + \frac{\psi}{6} \Pi^2 \right)
    + \frac{\sqrt{q}}{2} \bigg( \psi \rho - \psi R - F \left( \rho \right) + 2 \Delta \psi \bigg).
        \label{eq:geo_constraint}
\end{equation}
There is a secondary constraint, $0\approx\psi-F^{\,\prime}\left(\rho\right)$, which can be inverted and used to remove $\rho$ from $C$ as defined above. This leaves us with a term depending on $\psi$ which acts like a scalar field potential,
\begin{equation}
    U_\mathrm{geo} \left( \psi \right) = \frac{\psi}{2} \left( F' \right)^{-1} \left( \psi \right) - \half F \left( \left( F' \right)^{-1} \left( \psi \right) \right),
        \label{eq:geo_potential}
\end{equation}
which we call the geometric scalar potential.

We use the structure of the resulting constraint \eqref{eq:geo_constraint} to guide the structure of our general ansatz,
\begin{align}
    C = C_\0 + C_{(R)} R + C_{(\psi_i'\psi_j')} \partial_a \psi_i \partial^a \psi_j + C_{(\psi_i'')} \Delta \psi_i + C^{(p^2)}_{abcd} p^{ab} p^{cd} + C^{(p\Pi_i)} p \Pi_i + C^{(\Pi_i\Pi_j)} \Pi_i \Pi_j.
        \label{eq:ansatz}
\end{align}
We have generalised to include multiple scalar fields and the third term has been included because it is known to appear in the constraint for minimally coupled scalar fields.
We aimed to define the most general ansatz for a scalar-tensor constraint containing up to two orders in derivatives which is covariant under general spatial diffeomorphisms, as well as under time reversal, and preserves spatial parity.  Each coefficient is potentially a function of $q$ and $\psi_i$, allowing for non-minimal coupling. The spatial indices of $C^{(p^2)}_{abcd}$ can only represent different combinations of the metric. The zeroth order term $C_\0$ might include terms such as scalar field potentials or perfect fluids and behaves as a generalised potential.

We find the general action for a non-minimally coupled, second order, deformed scalar-tensor model in \secref{sec:action}, and extend this to multiple scalar fields in \secref{sec:multiple}.  We then turn to investigate the cosmological dynamics in \secref{sec:cosmo} and summarise our results in \secref{sec:summary}.
Appendix \ref{sec:time-asym} contains a discussion about time-reversal symmetry violating terms, and appendix \ref{sec:glossary} has a glossary of useful definitions.


\section{Constraining the action}
\label{sec:action}

Having our ansatz for the Hamiltonian constraint \eqref{eq:ansatz}, we need to find conditions for the coefficients to ensure that the constraint satisfies the correct symmetries. The Hamiltonian constraint and the diffeomorphism constraint, $\textstyle{}D_a:=\funcdif{H}{N^a}$, form an algebra (technically a Lie algebroid \cite{Bojowald:2016hgh}) which encodes the space-time diffeomorphism invariance,
\begin{subequations}
\begin{align}
    \big\{ D [N^a], D [M^b] \big\} & = D \big[ \mathcal{L}_M N^a \big],
        \label{eq:constraint-algebra_DD}
        \\
    \big\{ C [N], D [M^a] \big\} & = C \big[ \mathcal{L}_M N \big],
        \label{eq:constraint-algebra_CD}
        \\
    \big\{ C [N], C [M] \big\} & = D \big[ \beta \, q^{ab} \left( N \partial_b M - \partial_b N M \right) \big],
        \label{eq:constraint-algebra_CC}
\end{align}
    \label{eq:constraint-algebra}%
\end{subequations}
where the Poisson brackets are defined in \eqref{eq:poisson}.
We have included the deformation function $\beta$, which is a scalar dependent on phase space variables which deforms the general covariance, and $\beta\to1$ in the classical limit.  As stated before, we include the deformation to allow semi-classical corrections of the form suggested by loop quantum cosmology \cite{Cailleteau2012a, Mielczarek:2012pf}.

For a chosen field content, we can determine the unique form of the diffeomorphism constraint from \eqref{eq:constraint-algebra_DD}.  For a metric tensor field
$\textstyle\left(q_{ab},p^{cd}\right)$
and a scalar field
$\textstyle\left(\psi,\Pi\right)$,
this is given by,
\begin{equation}
    D_a = \nabla_a \psi \, \Pi - 2 q_{ab} \nabla_c p^{bc}.
        \label{eq:diffeomorphism_constraint}
\end{equation}
For the deformation function to affect the Hamiltonian constraint and not the diffeomorphism constraint, it must be a weightless scalar and the Hamiltonian constraint must be a scalar density of weight one.

Taking \eqref{eq:constraint-algebra_CC} and considering functional derivatives with respect to the functions $N(x)$ and $M(y)$, we find the distribution equation form,
\begin{equation}
    0 = \int \mathrm{d}^3 z \left(
        \funcdif{C\left(x\right)}{q_{ab}\left(z\right)} \funcdif{C\left(y\right)}{p^{ab}\left(z\right)}
        + \funcdif{C\left(x\right)}{\psi\left(z\right)} \funcdif{C\left(y\right)}{\Pi\left(z\right)}
    \right)
    - \beta (x) D^a (x) \partial_a \delta \left( x, y \right)
    - \left( x \leftrightarrow y \right).
        \label{eq:dist-eqn}
\end{equation}
Substituting into this the diffeomorphism constraint \eqref{eq:diffeomorphism_constraint}, the ansatz for $C$ \eqref{eq:ansatz}, and finding a form for $\beta$, we can then constrain the form of $C$.  Hence we must now choose the form which $\beta$ should take.

For general metrics, the distribution equation is linearly independent for different orders of momenta.  Considering only scalar variables, we collect the terms which are of order $\Pi^n$,
\begin{equation}
    0 = \sum_m \int \mathrm{d}^3 z \funcdif{C^{(m)}(x)}{\psi(z)} \funcdif{C^{(n-m+1)}(y)}{\Pi(z)} - \beta^{(n-1)}(x) D^a (x) \partial_a \delta \left( x , y \right) - \left( x \leftrightarrow y \right) ,
        \label{eq:dist-eqn_orders}
\end{equation}
where $C^{(n)}\sim\Pi^n$.  We assume that the highest order in momenta in the constraint, $n_C$, is the same as the highest order in spatial derivatives so the highest order terms are of the form $\Pi^{n_C-m}\partial\psi^{m}$ or $\Pi^{n_C-m}\Delta\psi^{m/2}$, where $m$ must be an even integer unless spatial parity is broken.  This means we can find the highest $n$ for which the first term in \eqref{eq:dist-eqn_orders} does not vanish under the $\left(x\leftrightarrow{}y\right)$ symmetry,
\begin{equation}
    0 = \int \mathrm{d}^3 z \funcdif{C^{(n_C-2)}(x)}{\psi(z)} \funcdif{C^{(n_C)}(y)}{\Pi(z)} - \beta^{(n-1)} (x) D^a (x) \partial_a \delta \left( x , y \right) - \left( x \leftrightarrow y \right),
        \quad
    n = 2 n_C - 3,
        \label{eq:dist-eqn_highest}
\end{equation}
and so the highest order terms in $\beta$ that need to be considered for a constraint of order $n_C$ are given by $n_\beta=2\left(n_C-2\right)$.  This result is not changed if we include metric variables. However, if we relax time symmetry and spatial parity, then $n_\beta=2n_C-3$.
Therefore, for a deformed second order constraint which conserves those symmetries, we only need to consider a zeroth order deformation $\beta=\beta\left(q,\psi\right)$ and likewise for a fourth order constraint we must consider a fourth order deformation.

We substitute into our distribution equation \eqref{eq:dist-eqn} our ansatz for a second order constraint \eqref{eq:ansatz}, the diffeomorphism constraint \eqref{eq:diffeomorphism_constraint}, and a zeroth order deformation $\beta\left(q,\psi\right)$,
\begin{equation}
\begin{split}
    0 & = \funcdif{C_0(x)}{q_{ab}(y)} \left( 2 p^{cd} C^{(p^2)}_{abcd} + \Pi q_{ab} C^{(p\Pi)} \right)_y + \funcdif{C_0(x)}{\psi(y)} \left( p \, C^{(p\Pi)} + 2 \Pi \, C^{(\Pi^2)} \right)_y
        \\
    & \quad + \left\{ 2 \beta \left( \partial_b p^{ab} + \Gamma^a_{bc} p^{bc} \right) - \beta \partial^a \psi \, \Pi \right\}_x \partial_a \delta(x,y) - \left( x \leftrightarrow y \right),
\end{split}
    \label{eq:def-st_dist-eqn}
\end{equation}
where $C_0$ is the part of the constraint without momenta, as defined in \eqref{eq:zero_subscript}.  From here there are two routes to solution, by focusing on either the $p^{ab}$ and $\Pi$ components and solving using the method outlined in \secref{sec:dist-eqn_method-1}.  We must do both to find all consistency conditions on the coefficients of the Hamiltonian constraint.

\subsection{\texorpdfstring{$p^{ab}$}{Metric momentum} sector}
\label{sec:def-st_p}

To proceed to the metric momentum sector, we take \eqref{eq:def-st_dist-eqn} and find the functional derivative with respect to $p^{ab}(z)$,
\begin{equation}
\begin{split}
    0 & = \left( 2 \funcdif{C_0(x)}{q_{cd}(y)} C^{(p^2)}_{abcd} (y) + \funcdif{C_0(x)}{\psi(y)} C^{(p\Pi)}_{ab} (y) \right) \delta(z,y)
        \\
    & \quad
    + 2 \beta (x) \left[ \delta^{cd}_{ab} \partial_{d(x)} \delta(z,x) + \Gamma^c_{ab}(x) \delta(z,x) \right] \partial_c \delta(x,y) - \left( x \leftrightarrow y \right),
\end{split}
    \label{eq:def-st_dist-eqn_p}
\end{equation}
where $\delta^{ab}_{cd}:=\delta^{(a}_c\delta^{b)}_d$ as in \eqref{eq:metric_combinations} and we have explicitly shown the partial derivative acting on the second variable of the delta function, rather than the first as is usual elsewhere.  This distinction is important when integrating by parts.
We then proceed by moving derivatives away from $\delta(z,y)$ terms and discarding total derivatives,
\begin{equation}
    0 = \left( 2 \funcdif{C_0(x)}{q_{cd}(y)} C^{(p^2)}_{abcd} (y) + \funcdif{C_0(x)}{\psi(y)} C^{(p\Pi)}_{ab} (y)
    + 2 \partial_{d} \left[ \beta(y) \delta^{cd}_{ab} \partial_{c} (y,x) \right] - 2 \beta(y) \Gamma^c_{ab}(y) \partial_c \delta (y,x) \right) \delta(z,y) - \left( x \leftrightarrow y \right),
\end{equation}
which we can rewrite as,
\begin{equation}
    0 = X_{ab}(x,y) \delta(z,y) - X_{ab}(y,x) \delta(z,x).
\end{equation}
Integrating over $y$, we find that part of the equation can be combined into a tensor dependent only on $x$,
\begin{equation}
\begin{split}
    0 & = X_{ab}(x,z) - \delta(z,x) \int \mathrm{d}^3 y X_{ab}(y,x),
        \\
    & = X_{ab}(x,z) - \delta(z,x) X_{ab} (x),
    \quad \mathrm{where} \;
    X_{ab}(x) = \int \mathrm{d}^3 y X_{ab} \left(y, x \right).
\end{split}
\end{equation}
Substituting in the definition of $X_{ab}(x,z)$ then relabelling,
\begin{equation}
\begin{split}
    0 & = 2 \funcdif{C_0(x)}{q_{cd}(y)} C^{(p^2)}_{abcd} (y) + \funcdif{C_0(x)}{\psi(y)} C^{(p\Pi)}_{ab} (y) + 2 \partial_{d} \left[ \beta(y) \delta^{cd}_{ab} \partial_{c} (y,x) \right]
        \\
    & \quad - 2 \beta(y) \Gamma^c_{ab}(y) \partial_c \delta (y,x) - X_{ab}(x) \delta(y,x).
\end{split}
\end{equation}
Multiplying by an arbitrary test tensor $\theta^{ab}\left(y\right)$, then integrating by parts over $y$, we get
\begin{equation}
\begin{split}
    0 & = \theta^{ab} \left( \cdots \right)_{ab}
    + \partial_c \theta^{ab} \left\{ 2 C^{(p^2)}_{abde} \partdif{C_0}{q_{de,c}} + 4 \partial_d C^{(p^2)}_{abef} \partdif{C_0}{q_{ef,cd}} + C^{(p\Pi)}_{ab} \partdif{C_0}{\psi_{,c}} + 2 \partial_d C^{(p\Pi)}_{ab} \partdif{C_0}{\psi_{,cd}} + 2 \delta^c_{(a} \partial_{b)} \beta + 2 \beta \Gamma^c_{ab} \right\}
        \\
    & \quad
    + \partial_{cd} \theta^{ab} \left\{ 2 C^{(p^2)}_{abef} \partdif{C_0}{q_{ef,cd}} + C^{(p\Pi)}_{ab} \partdif{C_0}{\psi_{,cd}} + 2 \beta \delta^{cd}_{ab} \right\},
\end{split}
\end{equation}
where we do not need to consider the zeroth derivative terms.  Since $\theta^{ab}$ is arbitrary, each unique combination of it forms a linearly independent equation.

To calculate the derivatives of $C_0$, we must use the decomposition of the Riemann tensor \eqref{eq:var_riemann_2} and the second covariant derivative of the metric variation expressed in terms of partial derivatives \eqref{eq:var_metric}.  This gives,
\begin{equation}
\begin{gathered}
    \partdif{C_0}{\psi_{,ab}}
    = C_{(\psi'')} q^{ab},
        \quad
    \partdif{C_0}{\psi_{,a}}
    = 2 C_{(\psi^{\prime2})} \partial^a \psi - C_{(\psi'')} \Gamma^a,
        \quad
    \partdif{C_0}{q_{ab,cd}}
    = C_{(R)} \Phi^{abcd},
        \\
    \partdif{C_0}{q_{ab,c}}
    = C_{(\psi'')} \left( \half q^{ab} \partial^c \psi - q^{c(a} \partial^{b)} \psi \right) - C_{(R)} \Phi^{defg} \left( \Gamma^c_{fg} \delta^{ab}_{de} + 4 \delta^{(a}_{(d} \Gamma^{b)}_{e)(f} \delta^c_{g)} \right),
\end{gathered}
    \label{eq:C0_derivatives}%
\end{equation}
where $\Phi^{abcd}$ is given by \eqref{eq:var_coeff_contract}.  We evaluate the coefficient of $\partial_{dc}\theta^{ab}$ and find the linearly independent components,
\begin{subequations}
\begin{align}
    q_{ab} \partial^2 \theta^{ab} :
    0 & = - 2 C_{(R)} \left( 2 C^{(p^2\p)} + C^{(p^2\x)} \right) + C_{(\psi'')} C^{(p\Pi)},
        \label{eq:d2epsilon_1} \\
    \partial_{ab} \theta^{ab} :
    0 & = C_{(R)} C^{(p^2\x)} + \beta,
        \label{eq:d2epsilon_2}
\end{align}
    \label{eq:d2epsilon}%
\end{subequations}
where we have decomposed the constraint coefficient
$C^{(p^2)}_{abcd} = q_{ab} q_{cd} C^{(p^2\p)} + Q_{abcd} C^{(p^2\x)}$.
Then evaluate similarly for $\partial_c\theta^{ab}$,
\begin{subequations}
\begin{align}
\begin{split}
    q_{ab} \partial_c \theta^{ab} \partial^c \psi :
    0 & = 2 \left( C_{(\psi^{\prime2})} + C_{(\psi'')} \partial_\psi \right) C^{(p\Pi)}
    + \left( C_{(\psi'')} - 8 C_{(R)} \partial_\psi \right) C^{(p^2\p)}
    + \left( C_{(\psi'')} - 4 C_{(R)} \partial_\psi \right) C^{(p^2\x)},
        \label{eq:d1epsilon_1} 
\end{split}
        \\
    \partial_a \theta^{ab} \partial_b \psi :
    0 & = \left( - C_{(\psi'')} + 2 C_{(R)} \partial_\psi \right) C^{(p^2\x)} + \partial_\psi \beta,
        \label{eq:d1epsilon_2} \\
    \partial_a \theta^{ab} q^{cd} \partial_b q_{cd} :
    0 & = C_{(R)} \left( 1 + 2 \partial_q \right) C^{(p^2\x)} + \partial_q \beta,
        \label{eq:d1epsilon_5} \\
    q_{ab} \partial_c \theta^{ab} q^{de} \partial^c q_{de} :
    0 & = - 2 C_{(R)} \left( 1 + 4 \partial_q \right) \left( 2 C^{(p^2\p)} + C^{(p^2\x)} \right) + C_{(\psi'')} \left( 1 + 4 \partial_q \right) C^{(p\Pi)},
        \label{eq:d1epsilon_7}
\end{align}
    \label{eq:d1epsilon}%
\end{subequations}
where $\textstyle\partial_\psi:=\partdif{}{\psi}$ and $\textstyle\partial_q:=\partdif{}{\,\ln{q}}$.
Note that the equations for $\partial_c\theta^{ab}\partial^cq_{ab}$, $\partial^c\theta^{ab}\partial_aq_{bc}$ and $q_{ab}\partial^c\theta^{ab}\partial^dq_{cd}$ are not included because they are identical to \eqref{eq:d2epsilon}.

Using \eqref{eq:d2epsilon_2} to solve for $C^{(p^2\x)}$, then substituting it into \eqref{eq:d1epsilon_5}, we find,
\begin{equation}
    \partdif{\,\ln{C}_{(R)}}{\,\ln{q}} = \half \left( 1 + \partdif{\,\ln{\beta}}{\,\ln{q}} \right),
\end{equation}
which is solved by $C_{(R)}\left(q,\psi\right)=f\left(\psi\right)\sqrt{q\,\left|\beta\left(q,\psi\right)\right|}$, where $f(\psi)$ is some unknown function.  If we solve \eqref{eq:d2epsilon} for $C^{(p^2\p)}$ and $C^{(p^2\x)}$, then substitute them into \eqref{eq:d1epsilon_7}, we find a similar equation to the one above for $C_{(R)}$, and therefore $C_{(\psi'')}\left(q,\psi\right)=f_{(\psi'')}\left(\psi\right)\sqrt{q\,\left|\beta\left(q,\psi\right)\right|}$.  Taking \eqref{eq:d1epsilon_2} then substituting in for $C^{(p^2\x)}$, $C_{(R)}$ and $C_{(\psi'')}$, we find that $f_{(\psi'')}\left(\psi\right)=-2\partial_\psi f\left(\psi\right)$,
\begin{subequations}
\begin{align}
    C_{(R)} & = f \sqrt{ q \, \left| \beta \right| },
        &
    C_{(\psi'')} & = - 2 \partial_\psi f \sqrt{ q \, \left| \beta \right| }, 
        \label{eq:C0_sol} \\
    C^{(p^2\x)} & = \frac{-\sgn{\beta}}{f} \sqrt{ \frac{ \left| \beta \right| }{ q } },
        &
    C^{(p^2\p)} & = \frac{\sgn{\beta}}{2f} \sqrt{ \frac{ \left| \beta \right| }{ q } } - \frac{ \partial_\psi f }{ 2 f } C^{(p\Pi)},
        \label{eq:Cp2_sol}
\end{align}
    \label{eq:p_sector_sol}%
\end{subequations}
which is all the conditions which we can obtain from the metric momentum sector of the distribution equation.  The remaining conditions must be found in the scalar momentum sector.

\subsection{\texorpdfstring{$\Pi$}{Scalar momentum} sector}
\label{sec:def-st_pi}

Similar to \subsecref{sec:def-st_p} above, we take the functional derivative of \eqref{eq:def-st_dist-eqn} with respect to $\Pi(z)$,
\begin{equation}
\begin{split}
    0 & = \left( \funcdif{C_0(x)}{q_{ab}(y)} C^{(p\Pi)}_{ab} (y) + 2 \funcdif{C_0(x)}{\psi(y)} C^{(\Pi^2)} (y) \right) \delta(z,y)
    - \left( \beta \partial^a \psi \right)_x \delta(z,x) \partial_a \delta(x,y) - \left( x \leftrightarrow y \right),
\end{split}
    \label{eq:def-st_dist-eqn_pi}
\end{equation}
then exchange terms to find the coefficient of $\delta(z,y)$,
\begin{equation}
\begin{split}
    0 & = \left( \funcdif{C_0(x)}{q_{ab}(y)} C^{(p\Pi)}_{ab} (y) + 2 \funcdif{C_0(x)}{\psi(y)} C^{(\Pi^2)} (y) + \left( \beta \partial^a \psi \right)_y \partial_a \delta(y,x) \right) \delta(z,y) - \left( x \leftrightarrow y \right),
\end{split}
\end{equation}
which we can rewrite as,
\begin{subequations}
\begin{align}
    0 & = X (x,y) \delta(z,y) - X (y,x) \delta(z,x),
        \\
    0 & = X (x,z) - \delta(z,x) \int \mathrm{d}^3 y X (y,x),
        \\
    & = X (x,z) - \delta(z,x) X (x),
        \quad \mathrm{where} \;
    X (x) = \int \mathrm{d}^3 y X \left( y, x \right),
\end{align}%
\end{subequations}
leading to
\begin{equation}
\begin{split}
    0 = \funcdif{C_0(x)}{q_{ab}(y)} C^{(p\Pi)}_{ab} (y) + 2 \funcdif{C_0(x)}{\psi(y)} C^{(\Pi^2)} (y) + \left( \beta \partial^a \psi \right)_y \partial_a \delta(y,x) - X(x) \delta(y,x).
\end{split}%
\end{equation}
Multiplying by an arbitrary test function $\eta(y)$, then integrating by parts over $y$, we get
\begin{equation}
\begin{split}
    0 &= \eta \left( \cdots \right) + \partial_{ab} \eta \left( C^{(p\Pi)}_{cd} \partdif{C_0}{q_{cd,ab}} + 2 C^{(\Pi^2)} \partdif{C_0}{\psi_{,ab}} \right)
        \\
    & \quad
    + \partial_a \eta \left( C^{(p\Pi)}_{bc} \partdif{C_0}{q_{bc,a}} + 2 \partial_b C^{(p\Pi)}_{cd} \partdif{C_0}{q_{cd,ab}} + 2 C^{(\Pi^2)} \partdif{C_0}{\psi_{,a}} + 4 \partial_b C^{(\Pi^2)} \partdif{C_0}{\psi_{,ab}} - \beta \partial^a \psi \right).
\end{split}
\end{equation}
We then substitute in \eqref{eq:C0_derivatives} to find the linearly independent conditions,
\begin{subequations}
\begin{align}
    \partial^2 \eta :
    0 & = C_{(R)} C^{(p\Pi)} - C_{(\psi'')} C^{(\Pi^2)},
        \label{eq:dphi_1} \\
    \partial_a \eta \partial^a \psi :
    0 & = \left( \half C_{(\psi'')} - 4 C_{(R)} \partial_\psi \right) C^{(p\Pi)} + 4 \left( C_{(\psi^{\prime2})} + C_{(\psi'')} \partial_\psi \right) C^{(\Pi^2)} - \beta,
        \label{eq:dphi_2} \\
    \partial_a \eta q^{bc} \partial^a q_{bc} :
    0 & = C_{(R)} \left( 1 + 4 \partial_q \right) C^{(p\Pi)} - C_{(\psi'')} \left( 1 + 4 \partial_q \right) C^{(\Pi^2)}.
        \label{eq:dphi_3}
\end{align}
    \label{eq:dphi}%
\end{subequations}
Note that there is another condition from $\partial^a\eta\partial^bq_{ab}$, but it is identical to \eqref{eq:dphi_1}.

We can solve \eqref{eq:dphi_1} for $C^{(p\Pi)}=C_{(\psi'')}C^{(\Pi^2)}/C_{(R)}$, and then substitute into \eqref{eq:dphi_2} to find,
\begin{equation}
    0 = C^{(\Pi^2)} \left\{ C_{(\psi^{\prime2})} - \partial_\psi C_{(\psi'')} + \frac{C_{(\psi'')}}{C_{(R)}} \left( \partial_\psi C_{(R)} + \frac{C_{(\psi'')}}{8} \right) \right\} - \frac{\beta}{4},
\end{equation}
which we can solve for $C^{(\Pi^2)}$, and is the same conclusion we get from \eqref{eq:d1epsilon_1} (though we did not explicitly write it above because it is simpler to write it here).  The condition \eqref{eq:dphi_3} is solved when we substitute in all our results so far,
\begin{subequations}
\begin{align}
    C^{(\Pi^2)} & = \quarter \sgn{\beta} \sqrt{\frac{\left|\beta\right|}{q}} \left\{ \frac{C_{(\psi^{\prime2})}}{\sqrt{q\left|\beta\right|}} + 2 f'' - \frac{3f^{\prime2}}{2f} \right\}^{-1},
        \\
    C^{(p\Pi)} & = \frac{-f'}{2f} \sgn{\beta} \sqrt{\frac{\left|\beta\right|}{q}} \left\{ \frac{C_{(\psi^{\prime2})}}{\sqrt{q\left|\beta\right|}} + 2 f'' - \frac{3f^{\prime2}}{2f} \right\}^{-1},
\end{align}%
\end{subequations}
and if we collect all of our coefficients, we find the Hamiltonian constraint,
\begin{equation}
\begin{split}
    C & = \sqrt{q\left|\beta\right|} \Big( f R - 2 f' \Delta \psi  \Big) + C_{(\psi^{\prime2})} \partial_a \psi \partial^a \psi + \sqrt{q} \, U \left( q, \psi \right)
        \\
    & \quad + \sgn{\beta} \sqrt{\frac{\left|\beta\right|}{q}} \left\{ \frac{1}{f} \left( \frac{p^2}{2} - \tr{p^2} \right) + \frac{1}{4} \left( \Pi - \frac{f'}{f} p \right)^2 \left( \frac{C_{(\psi^{\prime2})}}{\sqrt{q\left|\beta\right|}} + 2 f'' - \frac{3f^{\prime2}}{2f} \right)^{-1} \right\},
\end{split}
\end{equation}
so the freedom in any $(3+1)$ dimensional scalar-tensor theory with time symmetry and deformed general covariance comes down to the choice of $f\left(\psi\right)$, $\beta\left(q,\psi\right)$, $C_{(\psi^{\prime2})}\left(q,\psi\right)$ and the general potential $U\left(q,\psi\right)$.
It is convenient to make a redefinition, $C_{(\psi^{\prime2})}=g\left(q,\psi\right)\sqrt{q\left|\beta\right|}$, where we have made the scalar weight and expected dependence on $\beta$ explicit.  It is worth remembering that this is an assumption, and that $g$ could be a function of $\beta$.

We find the effective Lagrangian associated with this Hamiltonian constraint by performing a Legendre transformation,
\begin{equation}
\begin{split}
    L & = N \sqrt{q\left|\beta\right|} \left\{ f \left( \frac{\mathcal{K}}{\beta} - R \right) + f' \left( \frac{\nu{}v}{\beta} + 2 \Delta \psi \right) + \left( g + 2 f'' \right)\frac{\nu^2}{\beta} - g \, \partial_a \psi \partial^a \psi - \frac{U}{\sqrt{\left|\beta\right|}} \right\}.
\end{split}%
\end{equation}
Integrating by parts at the level of the action does not affect the dynamics because it only eliminates boundary terms.  This allows us to find the effective form of the Lagrangian, with a space-time decomposition and without second order time derivatives.  We can also do this in the opposite direction to find the covariant form of the above effective Lagrangian,
\begin{equation}
    L_\mathrm{cov} = N \sqrt{q\left|\beta\right|} \left( - f \, {}^{(4,\beta)}\!R - \left( g + 2 f'' \right) \partial_\mu^{(4,\beta)} \psi \, \partial^{\mu}_{(4,\beta)} \psi \right) - N \sqrt{q} \, U,
\end{equation}
where the deformed four dimensional Ricci scalar and partial derivative are given by,
\begin{subequations}
\begin{gather}
    {}^{(4,\beta)}\!R = R + \frac{\sgn{\beta}}{\sqrt{\left|\beta\right|}} q^{ab} \mathcal{L}_n \left( \frac{v_{ab}}{\sqrt{\left|\beta\right|}} \right) + \frac{1}{4\beta} v^2 - \frac{3}{4\beta} \tr{v^2} - \frac{2 \Delta \left( \sqrt{\left|\beta\right|} \,  N\right)}{\sqrt{\left|\beta\right|} \, N},
        \\
    \partial_\mu^{(4,\beta)} \psi \, \partial^{\mu}_{(4,\beta)} \psi = \partial_a \psi \, \partial^a \psi - \frac{1}{\beta} \nu^2,
\end{gather}%
\end{subequations}
whereby we see that the deformation seems to have transformed the effective lapse function $N\to\sqrt{\left|\beta\right|}\,N$, and transformed the effective normalisation of the normal vector to $g_{\mu\nu}n^\mu{}n^{\nu}=-\sgn{\beta}$.  Here is where we see the effective signature change which comes from the deformation.

It is useful to take the Lagrangian in covariant form and use it to redefine our coupling functions so that minimal coupling is when the functions are equal to unity,
\begin{equation}
    L_\mathrm{cov} = \half N \sqrt{q\left|\beta\right|} \left( \omega_R ( \psi ) {}^{(4,\beta)}\!R - \omega_\psi ( q, \psi ) \partial_\mu^{(4,\beta)} \psi \partial^\mu_{(4,\beta)} \psi \right) - N \sqrt{q} \, U \left( q, \psi \right),
        \label{eq:lagrangian_covariant}
\end{equation}
so the effective forms of the constraint and Lagrangian are given by,
\begin{subequations}
\begin{align}
    L & = \half N \sqrt{ q \left| \beta \right| } \left\{ \omega_R \left( R - \frac{\mathcal{K}}{\beta} \right) - \omega_R' \left( \frac{\nu{}v}{\beta} + 2 \Delta \psi \right) + \frac{\omega_\psi\nu^2}{\beta} 
    - \left( \omega_\psi + 2 \omega_R'' \right) \partial_a \psi \partial^a \psi \right\} - N \sqrt{q} \, U,
    \label{eq:lagrangian_effective}
        \\
\begin{split}
    C & = \sqrt{q\left|\beta\right|} \left\{ \frac{2\sgn{\beta}}{q\omega_R} \left( \tr{p^2} - \frac{p^2}{2} \right) - \frac{\omega_R}{2} R + \frac{\sgn{\beta}}{2q} \left( \Pi - \frac{\omega_R'}{\omega_R} p \right)^2 \left( \omega_\psi + \frac{3\omega_R^{\prime2}}{2\omega_R} \right)^{-1} \right.
        \\
    & \quad \left. + \omega_R' \Delta \psi + \left( \frac{\omega_\psi}{2} + \omega_R'' \right) \partial_a \psi \partial^a \psi \right\} + \sqrt{q} \, U,
\end{split}
    \label{eq:constraint_effective}
\end{align}
    \label{eq:effective}%
\end{subequations}
which is our result its most useful form.

Since we have non-minimal coupling, we are working in the Jordan frame.  We can get to the Einstein frame by making a specific conformal transformation which absorbs the coupling $\omega_R$ by setting $q_{ab}=\omega_R\,\tilde{q}_{ab}$ and $N=\omega_R^{-1/2}\tilde{N}$,
\begin{equation}
    \tilde{L} = \half \tilde{N}\sqrt{\tilde{q}\left|\beta\right|}
    \left\{
        \left( \tilde{R} - \frac{\tilde{\mathcal{K}}}{\beta} \right) + \left( \frac{\omega_\psi}{\omega_R} + \frac{3\omega_R^{\prime2}}{2\omega_R^2} \right) \left( \frac{\tilde{\nu}^2}{\beta} - \tilde{q}^{ab} \partial_a \psi \partial_b \psi \right)
    \right\} - \tilde{N} \sqrt{\tilde{q}} \left( \frac{U}{\omega_R^2} \right),
        \label{eq:lagrangian_einstein}
\end{equation}
where variables with tildes are Einstein-frame quantities.
So the Einstein frame couplings are given by $\tilde{\omega}_R=1$, $\tilde{\omega}_\psi=\left(\omega_\psi\omega_R+3\omega_R^{\prime2}/2\right)/\omega_R^2$, and the potential by $\tilde{U}=U/\omega_R^2$.

When the term `Einstein frame' is used elsewhere in the literature, it often refers to an action which is transformed further so that the effective scalar coupling is also unity.  We can make this transformation to a conventional scalar $\tilde{\psi}$ by solving the differential equation,
\begin{equation}
    \partdif{\tilde{\psi}}{\psi} = \sqrt{\frac{\omega_\psi}{\omega_R} + \frac{3}{2} \left( \frac{\partial_\psi\omega_R}{\omega_R} \right)^2},
\end{equation}
for example, when we have $\omega_\psi=0$, this is solved by
$
\textstyle \tilde{\psi} \left( \psi \right) = \sqrt{ \frac{3}{2} } \ln{ \omega_R \left( \psi \right) } \sgn{ \partial_\psi \ln{ \omega_R \left( \psi \right) } }
$.


\section{Multiple scalar fields}
\label{sec:multiple}

Let us consider the case of multiple scalar fields.  We start from the distribution equation as before, but label the scalar field variables with an index.  Proceeding like in \secref{sec:def-st_p} by taking functional derivatives with respect to $p^{ab}$ and then integrating by parts with test function $\theta^{ab}$, we obtain the conditions,
\begin{subequations}
\begin{align}
    \partial_{ab} \theta^{ab} : 0 & =
    C_{(R)} C^{(p^2\x)} + \beta,
        \label{eq:multiple_conditions_p1} 
        \\
    q_{ab} \partial^2 \theta^{ab} : 0 & =
    - 2 C_{(R)} \left( 2 C^{(p^2\p)} + C^{(p^2\x)} \right) + \sum_I C_{(\psi_I'')} C^{(p\Pi_I)},
        \label{eq:multiple_conditions_p2} 
        \\
    \partial_a \theta^{ab} q^{cd} \partial_b q_{cd} : 0 & =
    C_{(R)} \left( 1 + 2 \partial_q \right) C^{(p^2\x)} + \partial_q \beta,
        \label{eq:multiple_conditions_p3}
        \\
    \partial_a \theta^{ab} \partial_b \psi_I : 0 & = 
    \left( C_{(\psi_I'')} - 2 C_{(R)} \partial_{\psi_I} \right) C^{(p^2\x)} - \partial_{\psi_I} \beta,
        \label{eq:multiple_conditions_p4} 
        \\
\begin{split}
    \partial_c \theta^{ab} q_{ab} \partial^c \psi_I : 0 & = 
    \left( C_{(\psi_I'')} - 8 C_{(R)} \partial_{\psi_I} \right) C^{(p^2\p)} + \left( C_{(\psi_I'')} - 4 C_{(R)} \partial_{\psi_I} \right) C^{(p^2\x)}
    + 2 \left( C_{(\psi_I^{\prime2})} + C_{(\psi_I'')} \partial_{\psi_I} \right) C^{(p\Pi_I)}
        \\
    & \quad + \sum_{J\neq{}I} \left( C_{(\psi_I'\psi_J')} + C_{(\psi_J'')} \partial_{\psi_I} \right) C^{(p\Pi_J)}.
\end{split}
        \label{eq:multiple_conditions_p9} 
\end{align}
    \label{eq:multiple_conditions_p}%
\end{subequations}
We note that there are other independent terms, but they do not produce any extra conditions.
Likewise, if we follow the route taken in \secref{sec:def-st_pi}, taking the functional derivative with respect to $\Pi_I$ then integrating by parts with test function $\eta_I$, we find the conditions,
\begin{subequations}
\begin{align}
    \partial^2 \eta_I :
    0 & = C_{(R)} C^{(p\Pi_I)} - C_{(\psi_I'')} C^{(\Pi_I^2)} - \half \sum_{J\neq{}I} C_{(\psi_J'')} C^{(\Pi_I\Pi_J)},
        \label{eq:multiple_conditions_pi1}
        \\
    \partial_a \eta_I q^{bc} \partial^a q_{bc} :
    0 & = C_{(R)} \left( 1 + 4 \partial_q \right) C^{(p\Pi_I)} - C_{(\psi_I'')} \left( 1 + 4 \partial_q \right) C^{(\Pi_I^2)}
    - \half \sum_{J\neq{}I} C_{(\psi_J'')} \left( 1 + 4 \partial_q \right) C^{(\Pi_I\Pi_J)},
        \label{eq:multiple_conditions_pi2}
        \\
\begin{split}
    \partial_a \eta_I \partial^a \psi_I :
    0 & = \left( \half C_{(\psi_I'')} - 4 C_{(R)} \partial_{\psi_I} \right) C^{(p\Pi_I)} + 4 \left( C_{(\psi_I^{\prime2})} + C_{(\psi_I'')} \partial_{\psi_I} \right) C^{(\Pi_I^2)}
        \\
    & \quad + \sum_{J\neq{}I} \left( C_{(\psi_I'\psi_J')} + 2 C_{(\psi_J'')} \partial_{\psi_I} \right) C^{(\Pi_I\Pi_J)} - \beta,
\end{split}
        \label{eq:multiple_conditions_pi3}
        \\
\begin{split}
    \partial_a \eta_I \partial^a \psi_{J\neq{}I} :
    0 & = \left( \half C_{(\psi_J'')} - 2 C_{(R)} \partial_{\psi_J} \right) C^{(p\Pi_I)} + 2 \left( C_{(\psi_I'\psi_J')} + 2 C_{(\psi_I'')} \partial_{\psi_J} \right) C^{(\Pi_I^2)}
        \\
    & \quad + 2 \left( C_{(\psi_J^{\prime2})} + C_{(\psi_J'')} \partial_{\psi_J} \right) C^{(\Pi_I\Pi_J)}
    + \sum_{K\neq{}I,J} \left( C_{(\psi_J'\psi_K')} + 2 C_{(\psi_K'')} \partial_J \right) C^{(\Pi_I\Pi_K)},
\end{split}
        \label{eq:multiple_conditions_pi4}
\end{align}
        \label{eq:multiple_conditions_pi}%
\end{subequations}
and similar to above, there are other independent terms which do no produce any unique conditions.

To solve this system of equations we must make assumptions, in particular about the relationship between the scalar fields.  One choice might be to assume an $O\left(N\right)$ symmetry, where the coupling and deformation would only depend on the absolute value of the scalar field multiplet $\textstyle\left|\psi\right|=\sqrt{\sum_I\psi_I^2}$, and relationships between the $C_{(\psi_I'\psi_J')}$ coefficients could be assumed.

However, we instead choose to take one non-minimally coupled field $\left(\psi,\Pi\right)$ and one minimally coupled field $\left(\phi,\pi\right)$ with no cross-terms in the spatial derivative sector, $C_{(\phi'\psi')}=0$.  The minimally coupled field only appears in terms other than the potential $U\left(q,\psi,\phi\right)$ through the deformation function $\beta(q,\psi,\phi)$.  For example, $C_{(R)}=C_{(R)}\left(q,\psi,\beta\right)$.

Solving \eqref{eq:multiple_conditions_p1} and \eqref{eq:multiple_conditions_p3} gives us,
\begin{equation}
    C_{(R)} = f \left( \psi \right) \sqrt{q \left| \beta \left( q, \psi, \phi \right) \right|},
        \quad
    C^{(p^2\x)} = \frac{-1}{f\left(\psi\right)} \sqrt{\frac{\left| \beta \left( q, \psi, \phi \right) \right|}{q}},
\end{equation}
as before.  Substituting these into \eqref{eq:multiple_conditions_p2} and \eqref{eq:multiple_conditions_p4} gives us,
\begin{equation}
    C_{(\psi'')} = - 2 f' \sqrt{q\left|\beta\right|},
        \quad
    C_{(\phi'')} = 0,
        \quad
    C^{(p^2\p)} = \frac{\sgn{\beta}}{2f} \sqrt{\frac{\left|\beta\right|}{q}} - \frac{f'}{2f} C^{(p\Pi)},
\end{equation}
and the remaining conditions are,
\begin{subequations}
\begin{gather}
    C^{(p\Pi)}  = 
    \frac{-f'}{2f} \sgn{\beta} \sqrt{\frac{\left|\beta\right|}{q}} 
    \left\{
        \frac{C_{(\psi^{\prime2})}}{\sqrt{q\left|\beta\right|}} + 2 f'' - \frac{3f^{\prime2}}{2f}
    \right\}^{-1}
        \\
    C^{(\pi\Pi)} = 
    \frac{-\partial_\phi\beta\,\partial_\psi{}f}{4C_{(\phi^{\prime2})}}
    \left\{
    \frac{
        \frac{2C_{(\psi^{\prime2})}}{\sqrt{q\left|\beta\right|}} \left( 1 - \partdif{\,\ln{}C_{(\psi^{\prime2})}}{\,\ln{}\beta} \right) + 2 f'' - \frac{3f^{\prime2}}{2f}
    }{
        \left[ \frac{C_{(\psi^{\prime2})}}{\sqrt{q\left|\beta\right|}} + 2 f'' - \frac{3f^{\prime2}}{2f} \right]^2
    }
    \right\},
        \\
    C^{(\pi^2)} = \frac{\beta}{4C_{(\phi^{\prime2})}},
        \quad
    C^{(p\pi)} =
        -\frac{f'}{f} C^{(\pi\Pi)}.
\end{gather}%
\end{subequations}
We note that the constraint is significantly simpler if we assume $C_{(\phi^{\prime2})}=g_\phi\left(\psi\right)\sqrt{q\left|\beta\right|}$ and $C_{(\psi^{\prime2})}=g_\psi\left(\psi\right)\sqrt{q\left|\beta\right|}$, where $g_\phi$ and $g_\psi$ are arbitrary functions.
In this case the whole Hamiltonian constraint is
\begin{equation}
\begin{split}
    C & = \sqrt{q\left|\beta\right|} \Big( f R - 2 f' \Delta \psi + g_\phi \partial_a \phi \partial^a \phi + g_\psi \partial_a \psi \partial^a \psi \Big) + \sqrt{q} \, U
        \\
    & \quad + \sgn{\beta} \sqrt{\frac{\left|\beta\right|}{q}} \left\{ \frac{\pi^2}{4g_\phi} + \frac{1}{f} \left( \frac{p^2}{2} - \tr{p^2} \right)
    + \frac{\left( \Pi - \frac{f'}{f} p \right) \left( \Pi - \frac{f'}{f} p - \frac{f'\partial_\phi\beta}{\beta{}g_\phi} \pi \right)
    }{4 \left( g_\psi + 2 f'' - \frac{3f^{\prime2}}{2f} \right)}
    \right\},
\end{split}
\end{equation}
and the associated Lagrangian density is
\begin{subequations}
\begin{gather}
\begin{split}
    L & = N \sqrt{q \left| \beta \right|} \left\{ f \left( \frac{\mathcal{K}}{\beta} - R \right) + f' \left( \frac{\nu_\psi v}{\beta} + 2 \Delta \psi \right) + \left( \frac{\hat{g}_\psi}{h} + \frac{3f^{\prime2}}{2f} \right) \frac{\nu_\psi^2}{\beta} - g_\psi \partial_a \psi \partial^a \psi \right.
        \\
    & \left. \quad + \frac{g_\phi}{h\beta} \nu_\phi^2 - g_\phi \partial_a \phi \partial^a \phi + \frac{f'\partial_\phi\beta}{h\beta} \nu_\phi \nu_\psi \right\} - N \sqrt{q} U,
\end{split}
        \\
    \hat{g}_{\psi} = g_\psi + 2 f'' - \frac{3f^{\prime2}}{2f},
        \quad \quad
    h = 1 - \frac{f^{\prime2}\partial_\phi\beta^2}{4g_\phi\hat{g}_\psi\beta^2}.
\end{gather}%
    \label{eq:multiple_lagrangian_rough}%
\end{subequations}
If $\beta$ does not depend on $\phi$, then this can be simplified greatly, in which case the effective and covariant forms of the Lagrangian are given by,
\begin{subequations}
\begin{align}
\begin{split}
    L & = \half N \sqrt{ q \left| \beta \right| } \left\{ \omega_R \left( R - \frac{\mathcal{K}}{\beta} \right) - \omega_R' \left( \frac{\nu_\psi{}v}{\beta} + 2 \Delta \psi \right) + \omega_\phi \bigg( \frac{\nu_\phi^2}{\beta} - \partial_a \phi \partial^a \phi \bigg) \right.
        \\
    & \quad \left.
    + \frac{\omega_\psi\nu_\psi^2}{\beta} - \left( \omega_\psi + 2 \omega_R'' \right) \partial_a \psi \partial^a \psi  \right\} - N \sqrt{q} \, U,
\end{split}
        \\
    L_\mathrm{cov} & = \half N \sqrt{q\left|\beta\right|} \left( \omega_R \, {}^{(4,\beta)}\!R - \omega_\psi \, \partial_\mu^{(4,\beta)} \psi \partial^\mu_{(4,\beta)} \psi - \omega_\phi \, \partial_\mu^{(4,\beta)} \phi \partial^\mu_{(4,\beta)} \phi \right) - N \sqrt{q} \, U,
\end{align}%
    \label{eq:multiple_lagrangian}%
\end{subequations}
where $\omega_R=-2f$, $\omega_\psi=2\left(g_\psi+2f''\right)$, $\omega_\phi=2g_\phi$.  Therefore, when we assume that the minimally coupled scalar field can also be considered to be minimally coupled to the deformation function, we find that the action simplifies to the expected form.
It would be interesting to see what effects appear for scalar field multiplets, especially for non-Abelian symmetries, but that is beyond the scope of this study.
Instead, we now turn to studying the cosmological dynamics of our results.


\section{Cosmology}
\label{sec:cosmo}

Let us restrict to a flat, homogeneous, and isotropic metric in proper time ($N=1$) to find the background dynamics.  We also assume that $\beta$ does not depend on the minimally coupled scalar field $\phi$ for the sake of simplicity.  From \eqref{eq:multiple_lagrangian}, we find the Friedmann equation, which can be written in two equivalent forms,
\begin{subequations}
\begin{align}
    H \left( \omega_R H + \omega_R' \dot{\psi} \right)
    & =
    \third \left( \frac{\omega_\psi}{2} \dot{\psi}^2 + \frac{\omega_\phi}{2} \dot{\phi}^2 + \sgn{\beta} \sqrt{\left|\beta\right|} \, U  \right),
        \label{eq:cosmo_friedmann_1}
        \\
    \left( \omega_R H + \half \omega_R' \dot{\psi} \right)^2
    & =
    \third \left[ \half \left( \omega_R \omega_\psi + \frac{3}{2} \omega_R^{\prime2} \right) \dot{\psi}^2 + \frac{\omega_R \omega_\phi}{2} \dot{\phi}^2 + \omega_R \sgn{\beta} \sqrt{\left|\beta\right|} \, U  \right].
        \label{eq:cosmo_friedmann_2}
\end{align}
    \label{eq:cosmo_friedmann}%
\end{subequations}
From \eqref{eq:cosmo_friedmann_2} we see that $\omega_R\omega_\psi+3\omega_R^{\prime2}/2\geq0$ and $\omega_R\omega_\phi\geq0$ are necessary when $U\to0$ to ensure real-valued fields.  If we compare this condition to the Einstein frame Lagrangian \eqref{eq:lagrangian_einstein}, we can see that it is also the condition which follows from insisting that the scalar field $\psi$ is not ghost-like in that frame.  Similarly, we see that $\omega_R\sgn{\beta}>0$ is necessary when $\dot{\psi},\dot{\phi}\to0$.

For the reasonable assumption that the minimally coupled field $\phi$ does not affect the deformation function $\beta$, the only way that field is modified is through a variable maximum phase speed $c_\phi^2=\beta$.  Due to this minimal modification, it does not produce any of the cosmological phenomena we are interested in (bounce, inflation) through any novel mechanism.  Therefore, we will ignore this field for the rest of the paper.

We find the equations of motion by varying the Lagrangian \eqref{eq:multiple_lagrangian} with respect to the fields.  
For the simple undeformed case $\beta=1$ the equations are given by,
\begin{subequations}
\begin{gather}
\begin{gathered}
    \left( \omega_R \omega_\psi + \frac{3}{2} \omega_R^{\prime2} \right) \ddot{\psi} =
    - 3 \dot{\psi} H \left( \omega_R \omega_\psi + \omega_R^{\prime2} \right) - \omega_R \partial_\psi U + \frac{3}{2} \omega_R \omega_R' H^2
        \\
    - \half \dot{\psi}^2 \left( \omega_R \omega_\psi' + \frac{3}{2} \omega_R' \omega_\psi + 3 \omega_R' \omega_R'' \right) + \frac{3}{2} \omega_R' \left( U + \third \partial_{\,\ln{a}} U \right),
\end{gathered}
    \label{eq:cosmo_scalar}
        \\
\begin{gathered}
    \left( \omega_R \omega_\psi + \frac{3}{2} \omega_R^{\prime2} \right) \frac{\ddot{a}}{a} =
    - \frac{1}{2} H^2 \left( \omega_R \omega_\psi + 3 \omega_R^{\prime2} \right) + \frac{\omega_\psi}{2} \left( U + \third \partial_{\,\ln{a}} U \right)
        \\
    - \quarter \dot{\psi}^2 \left( \omega_\psi^2 + 2 \omega_\psi \omega_R'' - \omega_\psi' \omega_R' \right) + \half \omega_R' \omega_\psi \dot{\psi} H - \frac{\omega_R'}{2} \partial_\psi U,
\end{gathered}
    \label{eq:cosmo_acceleration}
\end{gather}
    \label{eq:cosmo_eom}%
\end{subequations}
where we can see from the equations of motion that the model breaks down if $\omega_R\omega_\psi+3\omega_R^{\prime2}/2\to0$ because it will tend to cause $|\ddot{\psi}|\to\infty$ and $|\ddot{a}|\to\infty$.

\subsection{Bounce}
\label{sec:cosmo_bounce}

We will address the question of whether there are conditions under which a big bounce might occur (by which we mean a turning point preventing $a\to0$).
In a previous study we found that a deformation function which depends on curvature terms can generate a bounce \cite{Cuttell2014}. Elsewhere in the literature on loop quantum cosmology the bounce happens in a regime when $\beta<0$ because the terms depending on curvature or energy density overpower the zeroth order terms \cite{Cailleteau2012a, Mielczarek:2012pf}.  However, we are not including derivatives in the deformation here so the effect would have to come from the non-minimal coupling of the scalar field or the zeroth order deformation.

We take $\dot{a}=0$ for finite $a$, include a deformation and for simplicity we ignore the minimally coupled field.  From the Friedmann equation \eqref{eq:cosmo_friedmann} we find,
\begin{equation}
    0 = \frac{\omega_\psi}{2} \dot{\psi}^2 + \sgn{\beta} \sqrt{ \left| \beta \right| } \, U,
        \label{eq:cosmo_bounce}
\end{equation}
which implies that $\omega_\psi\sgn{\beta}<0$ for a bounce because otherwise the equation cannot balance for $U>0$ and $\psi\in\mathbb{R}$.
Substituting \eqref{eq:cosmo_bounce} into the full equation of motion for the scale factor, and demanding that $\ddot{a}>0$ to make it a turning point, we find the following conditions,
\begin{subequations}%
\begin{gather}
    \sgn{\beta} \omega_\psi < 0,
        \\
    \omega_R \omega_\psi + \frac{3}{2} \omega_R^{\prime2} > 0,
        \\
    \sgn{\beta} \sqrt{ \left| \beta \right| } \left( \omega_\psi + 2 \omega_R'' \right) U - \frac{\sgn{\beta}\omega_R'}{2\omega_\psi} \partial_\psi \left( \sqrt{ \left| \beta \right| } \omega_\psi U \right) + \frac{\beta}{6} \partial_{\,\ln{a}} \left( \frac{ \omega_\psi U }{ \sqrt{ \left| \beta \right| } } \right) > 0,
\end{gather}
    \label{eq:cosmo_bounce_conditions}%
\end{subequations}
from which we can determine what the coupling functions, deformation and potential must be for a bounce.
For example, if we look at the minimally coupled case, when $\omega_R=\omega_\psi=1$, and assume that $U>0$, we can see that the conditions are given by,
\begin{equation}
    \sgn{\beta} < 0,
        \quad
    \partdif{ \, \ln \left( \left| \beta \right|^{-1/2} U \right) }{ \, \ln{a} } < -6.
\end{equation}
Since we must have $\beta\to1$ in the classical limit and $\sgn{\beta}<0$ at the moment of the bounce, then $\beta$ must change sign at some point.  Therefore, a universe which bounces purely due to a zeroth order deformation must have effective signature change.
Another example is obtained by assuming scale independence and choosing $\beta=1$ and $U>0$.  In this case the bounce conditions become,
\begin{equation}
    \omega_\psi < 0,
        \quad
    \omega_\psi \omega_R + \frac{3}{2} \omega_R^{\prime2} > 0,
        \quad
    \omega_\psi + 2 \omega_R'' - \half \omega_R' \partial_\psi \ln \left( \omega_\psi U \right) > 0,
\end{equation}
which we can use to find a model which bounces purely due to a scale-independent non-minimally coupled scalar.  We present this model in \subsecref{sec:cosmo_BS}.

\subsection{Inflation}
\label{sec:cosmo_inflation}

Now we look at the inflationary dynamics.
For simplicity we assume that inflation will come from a scenario similar to slow-roll inflation with possible enhancements coming from the non-minimal coupling or the deformation.
The conditions for slow-roll inflation are,
\begin{equation}
    \dot{\psi}^2 \ll U, 
        \quad
    \left| \ddot{\psi} \right| \ll \left| \dot{\psi} H \right|,
        \quad
    \left| \dot{H} \right| \ll H^2,
        \label{eq:slow-roll_conditions}
\end{equation}
assuming the couplings, potential and deformation are scale independent and the deformation is positive, we get the following slow roll equations,
\begin{subequations}
\begin{align}
    H & \simeq
    \sqrt{ \frac{ \beta^{1/2} U }{ 3 \omega_R } },
        \\
    \dot{\psi} & \simeq
    - \sqrt{ \frac{ \beta^{1/2} U }{ 3 \omega_R } }
    \left( \frac{ \partial_\psi \ln \left( \frac{U}{ \beta^{1/2} \omega_R^2 } \right) }{ \frac{\omega_\psi}{\omega_R} + \frac{\omega_R^{\prime2}}{\omega_R^2} + \frac{\beta'\omega_R'}{2\beta\omega_R} } \right),
\end{align}%
    \label{eq:slow-roll_equations}%
\end{subequations}
and define the slow-roll parameters,
\begin{equation}
    \epsilon : = \frac{-\dot{H}}{H^2},
        \quad
    \eta : = \frac{-\ddot{H}}{H\dot{H}},
        \quad
    \zeta : = \frac{-\ddot{\psi}}{\dot{\psi}H},
        \label{eq:slow-roll_parameters_definition}
\end{equation}
which, under slow-roll conditions are given by,
\begin{subequations}
\begin{align}
    \epsilon & \simeq \frac{ \partial_\psi \ln \left( \frac{\beta^{1/2}U}{\omega_R} \right) \partial_\psi \ln \left( \frac{U}{\beta^{1/2}\omega_R^2} \right) }{ 2 \left( \frac{\omega_\psi}{\omega_R} + \frac{\omega_R^{\prime2}}{\omega_R^2} + \frac{\beta'\omega_R'}{2\beta\omega_R}  \right) }
        \label{eq:slow-roll_epsilon} \\
    \eta & \simeq \left( \frac{ \partial_\psi \ln \left( \frac{U}{\beta^{1/2}\omega_R^2} \right) }{ \frac{\omega_\psi}{\omega_R} + \frac{\omega_R^{\prime2}}{\omega_R^2} + \frac{\beta'\omega_R'}{2\beta\omega_R} } \right) \partial_\psi \ln{\epsilon} + 2 \epsilon,
        \label{eq:slow-roll_eta} \\
    \zeta & \simeq
    \partial_\psi \left(
    \frac{\partial_\psi \ln \left( \frac{U}{\beta^{1/2}\omega_R^2} \right) }{ \frac{\omega_\psi}{\omega_R} + \frac{\omega_R^{\prime2}}{\omega_R^2} + \frac{\beta'\omega_R'}{2\beta\omega_R} }
    \right)
    + \epsilon,
        \label{eq:slow-roll_zeta}%
\end{align}%
        \label{eq:slow-roll_parameters}%
\end{subequations}
and the slow-roll regime ends when the absolute value of any of these three parameters approaches unity.

Defining $\mathcal{N}$ to mean the number of e-folds from the end of inflation, $a\left(t\right)=a_\mathrm{end}e^{-\mathcal{N}\left(t\right)}$, we find that,
\begin{equation}
    \mathcal{N} = - \int_{t_\mathrm{end}}^t \mathrm{d} t H = - \int_{\psi_\mathrm{end}}^\psi \mathrm{d} \psi \frac{H}{\dot{\psi}},
        \label{eq:efolds_general}
\end{equation}
and using the slow-roll approximation,
\begin{equation}
    \mathcal{N} \simeq \int_{\psi_\mathrm{end}}^\psi \frac{ \frac{\omega_\psi}{\omega_R} + \frac{\omega_R^{\prime2}}{\omega_R^2} + \frac{\beta'\omega_R'}{2\beta\omega_R}  }{\partial_\psi \ln \left( \frac{ U }{ \beta^{1/2} \omega_R^2} \right) },
        \label{eq:efolds}
\end{equation}
which can be solved once we specify the form of the couplings, deformation and potential.
We cannot find equations for observables such as the spectral index $n_s$ because it would require investigating how the cosmological perturbation theory is modified in the presence of non-minimal coupling and deformed general covariance which is not within the scope of this study.
Beyond this, it is difficult to make general statements about the dynamics unless we restrict to a given model, so we will now consider some models and discuss their specific dynamics.

\subsection{Geometric scalar model}
\label{sec:cosmo_geo}

As demonstrated in the introduction, \secref{sec:intro}, the geometric scalar model comes from parameterising $\textstyle{}F\left({}^{(4)}\!R\right)$ gravity so that the additional degree of freedom of the scalar curvature is instead embodied in a non-minimally coupled scalar field $\psi$ \cite{Deruelle:2009pu, Deruelle2010}.
Its couplings are given by $\omega_R=\psi$ and $\omega_\psi=0$.  This model is a special case of the Brans-Dicke model, which has $\omega_\psi=\omega_0/\psi$, when the Dicke coupling constant $\omega_0$ vanishes.
We can add in a minimally coupled scalar field with $\omega_\phi=1$ and thereby see the effect of this scalar-tensor gravity on the matter sector.  However, we set $\omega_\phi=0$ because it does not significantly affect our results.

The effective action for this model is given by,
\begin{subequations}
\begin{gather}
\begin{split}
    L_\mathrm{geo} & = \half N \sqrt{q\left|\beta\right|} \left\{ \psi \left( R - \frac{\mathcal{K}}{\beta} \right) - \frac{\nu_\psi v}{\beta} - 2 \Delta \psi \right\} - N \sqrt{q} \, U \left( \psi \right),
\end{split}
        \\
    U \left( \psi \right) = \frac{\psi}{2} \left( F^{\,\prime} \right)^{-1} \left( \psi \right) - \half F \left( \left( F^{\,\prime} \right)^{-1} \left( \psi \right) \right),
\end{gather}%
\end{subequations}
where $F$ refers to the $\textstyle{}F\left(\Rfour\right)$ function which has been parameterised.  The equations of motion when $\beta\to1$ are given by,
\begin{subequations}
\begin{gather}
    H \left( \psi H + \dot{\psi} \right) 
    = \third U ,
        \\
    \frac{\ddot{a}}{a} 
    = - H^2  + \frac{1}{3} \partial_\psi U,
        \\
    \ddot{\psi} 
    = -2 \dot{\psi} H + \psi H^2 + U + \third \partial_{\,\ln{a}} U - \frac{2}{3} \partial_{\,\ln{\psi}} U,
\end{gather}%
\end{subequations}
from which we can see that the scalar field has very different dynamics compared to minimally coupled scalars.  This reflects its origin as a geometric degree of freedom rather than a purely matter field.

\begin{figure}[t]
	\begin{center}
	{\subfigure[Scale factor (Logarithmic)]{
	    \label{fig:geo_scale-factor-log}
		\includegraphics[height=3.3cm]{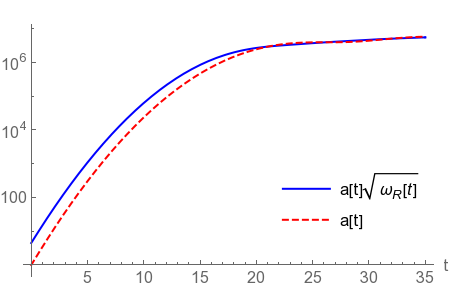}}}
	{\subfigure[Scale factor]{
		\label{fig:geo_scale-factor}
		\includegraphics[height=3.3cm]{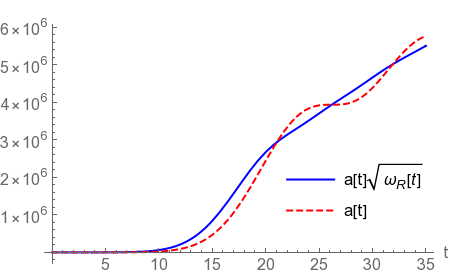}}}
	{\subfigure[Scalar field]{
	    \label{fig:geo_scalar}
		\includegraphics[height=3.3cm]{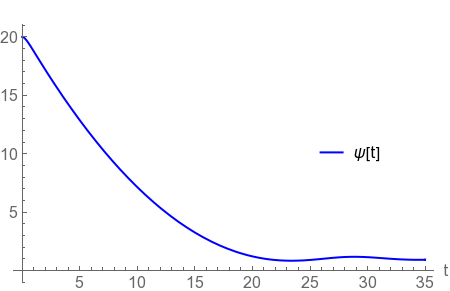}}}
	\end{center}
	\caption{Inflation from the geometric scalar model version of the Starobinsky model through slow-roll of the non-minimally coupled scalar field.  For the scale factor, we compare the Jordan and Einstein frames because the coupling causes the former to oscillate unusually.  Initial conditions, $a=1$, $\psi=20$, $\dot{\psi}=0$, $M=1$.}
    	\label{fig:geo}
\end{figure}

Looking at inflation, the geometric scalar model with a potential corresponding to the Starobinsky model,
\begin{equation}
    F \left( \Rfour \right) = \Rfour + \frac{1}{2M^2} \Rfour^2
    \quad \to \quad
    U = \frac{M^2}{4} \left( \psi - 1 \right)^2,
\end{equation}
can indeed cause inflation through a slow-roll of the scalar field down its potential.  The non-minimal coupling of the scalar to the metric also causes the scale factor to oscillate unusually, however.  It is interesting to compare in \figref{fig:geo} the scale factor in the Jordan frame, $a$, and the conformally transformed scale factor in the Einstein frame, $\tilde{a}=a\sqrt{\omega_R}$.
Assuming $\psi>1$ during inflation, the slow-roll parameters \eqref{eq:slow-roll_parameters} are given by,
\begin{equation}
    \epsilon \simeq
    \frac{\psi+1}{\left(\psi-1\right)^2},
    \quad
    \eta \simeq
    \frac{-2}{\psi^2-1},
    \quad
    \zeta \simeq
    \frac{1}{\psi-1},
\end{equation}
so the slow-roll regime of inflation ends at $\psi\approx3$ when $\epsilon\to1$.
The equation for the number of e-folds of inflation in the slow-roll regime \eqref{eq:efolds} is given by
$
\textstyle
    \mathcal{N} \simeq \half \left( \psi - \psi_\mathrm{end} - \ln{\frac{\psi}{\psi_\mathrm{end}}} \right)
$.

\subsection{Non-minimally enhanced scalar model}
\label{sec:cosmo_nes}

\begin{figure}[t]
	\begin{center}
	{\subfigure[]{
	    \label{fig:NES-exclusion}
		\includegraphics[height=0.333\textwidth]{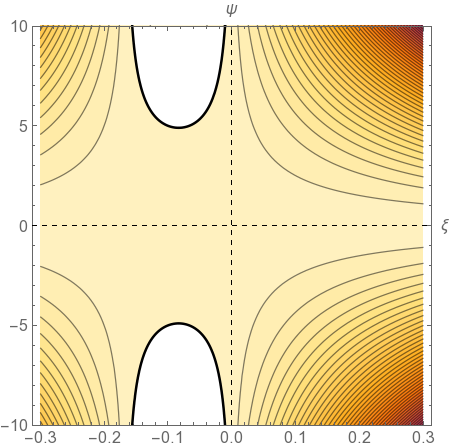}}}
	{\subfigure[]{
	    \label{fig:NES-ghostzones}
	    \includegraphics[height=0.333\textwidth]{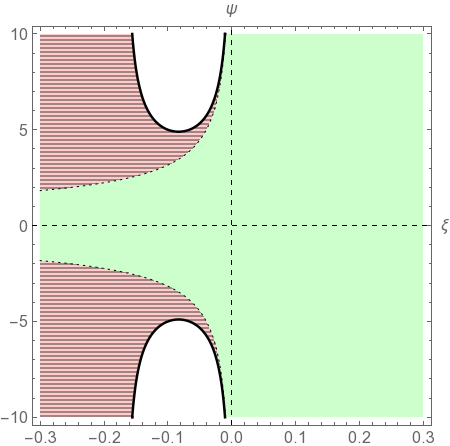}}}
	\end{center}
	\caption{A contour plot of $\omega_R\omega_\psi+3\omega_R^{\prime2}/2$ for the non-minimally enhanced scalar model is shown in \subref{fig:NES-exclusion}, and the white regions are forbidden because the function is negative there. In \subref{fig:NES-ghostzones}, the red, striped region is when the metric becomes ghost-like (when $\omega_R<0$).}
	    \label{fig:NS}
\end{figure}

Unlike the geometric scalar model considered above, the non-minimally enhanced scalar model (NES) from \cite{Nozari:2010uu}, takes a scalar field from the matter sector and introduces a non-minimal coupling rather than extracting a degree of freedom from the gravity sector.  The coupling functions are given by
$\omega_R = 1 + \xi \psi^2$, $\omega_\psi=1$ and $\omega_\phi=0$.
The deformed effective Lagrangian for this model is given by,
\begin{equation}
\begin{split}
    L_\mathrm{NES} & = N \sqrt{q\left|\beta\right|} \left\{ \half \left( 1 + \xi \psi^2 \right) \left( R - \frac{\mathcal{K}}{\beta} \right) + \half \left( \frac{\nu_\psi^2}{\beta} - \partial_a \psi \partial^a \psi \right)
    \right.
        \\
    & \quad \left. 
    - 2 \xi \left( \frac{\psi \nu_\psi v}{2\beta} + \psi \Delta \psi + \partial_a \psi \partial^a \psi \right) \right\} - N \sqrt{q} \, U \left( \psi \right).
\end{split}%
\end{equation}
For some negative values of $\xi$, there are values of $\psi$ which are forbidden if we are to keep our variables real, shown in \figref{fig:NS}.

\begin{figure}[t]
    \begin{center}
    {\subfigure[]{
        \label{fig:nes_inflation_numerical}
            \includegraphics[height=5cm]{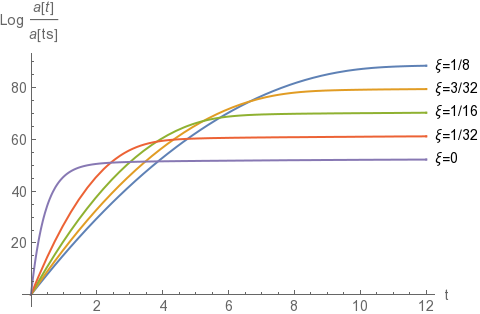}}}
    {\subfigure[]{
        \label{fig:nes_inflation_efolds}
            \includegraphics[height=5cm]{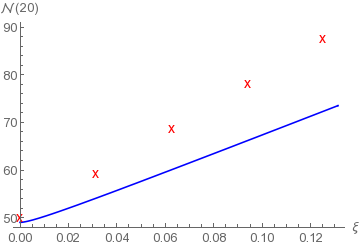}}}
    \end{center}
    \caption{For the non-minimally enhanced scalar model with $U=\psi^4/4$, \subref{fig:nes_inflation_numerical} shows numerical solutions of inflation for different coupling strengths. Initial conditions, $\psi=20$, $\dot{\psi}=0$, $H>0$.  In \subref{fig:nes_inflation_efolds}, $\mathcal{N}$ for $\psi=20$ is compared for the numerical solutions (red crosses) and the analytical solution in the slow-roll approximation \eqref{eq:efolds_NES} (blue line).}
    \label{fig:nes_inflation}
\end{figure}

The equations of motion for this model when it is undeformed are given by,
\begin{subequations}
\begin{gather}
    \left( 1 + \xi \psi^2 \right) H^2 + 2 \xi \psi \dot{\psi} H
    = \third \left( \half \dot{\psi^2} + U \right),
        \label{eq:cosmo_nes_friedmann} \\
\begin{split}
   \left( 1 + \left( 1 + 6 \xi \right) \xi \psi^2 \right) \frac{\ddot{a}}{a} 
    & = \frac{-1}{2} H^2 \left( 1 + \left( 1 + 12 \xi \right) \xi \psi^2 \right) - \frac{1+4\xi}{4} \dot{\psi}^2
        \\
    & \quad + \xi \psi \dot{\psi} H + \half \left( U + \third \partial_{\,\ln{a}} U \right) + \xi \partial_{\,\ln{\psi}} U,
\end{split}
        \label{eq:cosmo_nes_acceleration} \\
\begin{split}
    \left( 1 + \left( 1 + 6 \xi \right) \xi \psi^2 \right) \ddot{\psi}
    & = - 3 \dot{\psi} H \left( 1 + \left( 1 + 4 \xi \right) \xi \psi^2 \right) - \left( 1 + \xi \psi^2 \right) \partial_\psi U
        \\
    & \quad + 3 \xi \psi \left( \left( 1 + \xi \psi^2 \right) H^2 - \frac{1+4\xi}{2} \dot{\psi}^2 + U + \third \partial_{\,\ln{a}} U \right).
\end{split}
        \label{eq:cosmo_nes_scalar}
\end{gather}
    \label{eq:cosmo_ns}%
\end{subequations}
and we proceed to use them to consider this model's inflationary dynamics.  For $\xi>0$ the slow-roll parameter which reaches unity first is $\epsilon$ at $\textstyle \psi_\mathrm{end} \simeq \frac{\pm n}{\sqrt{ 2 + n \left( 6 - n \right) \xi }}$.  The number of e-folds from the end of inflation is given by,
\begin{equation}
    \mathcal{N}_\mathrm{NES} \left( \psi \right) \simeq \int^\psi_{\psi_\mathrm{end}} \mathrm{d} \varphi
    \frac{ \varphi \left( 1 + \left( 1 + 4 \xi \right) \xi \varphi^2 \right) }{ \left( 1 + \xi \varphi^2 \right) \left( n + \left( n - 4 \right) \xi \varphi^2 \right)},
\end{equation}
and if we specify that $n=4$, we find
\begin{equation}
    \mathcal{N}_\mathrm{NES} \simeq \frac{1+4\xi}{8} \psi^2 - 1 + \half \ln{\frac{1+12\xi}{\left(1+4\xi\right)\left(1+\xi\psi^2\right)}},
        \label{eq:efolds_NES}
\end{equation}
and the presence of $\xi$ in the dominant first term shows how the non-minimal coupling enhances the amount of inflation.  If we compare this result to numerical solutions in \figref{fig:nes_inflation}, we see this effect.

The slow-roll approximation works less well as $\xi$ increases.  We can see this when we look at \figref{fig:nes_inflation_efolds} where we compare the slow-roll approximation to when we numerically determine the end of inflation, i.e. when $\epsilon=-\dot{H}/H^2=1$.

We must be wary when dealing with this model, because the coupling can produce an effective potential which is not bounded from below.  If we substitute the Friedmann equation \eqref{eq:cosmo_nes_friedmann} into \eqref{eq:cosmo_nes_acceleration} and \eqref{eq:cosmo_nes_scalar} we can find effective potential terms. These terms are those which do not vanish when all time derivatives are set to zero, and we can infer what bare potential they effectively behave like.  If the bare potential is $U=\l\psi^2/2$, then the effective potential term in the scalar equation behaves like
\begin{equation}
    U_\psi = \frac{-\l\psi^2}{2\left(1+6\xi\right)} + \frac{\l \left( 1 + 3 \xi \right)}{\xi \left( 1 + 6 \xi \right)^2} \ln{\left( 1 + \left( 1 + 6 \xi \right) \xi \psi^2 \right)},
\end{equation}
which is not bounded from below when $\xi>0$ and $\lambda>0$ and is therefore unstable.  More generally, for a power law bare potential $U=\l\psi^n/n$, there are local maxima in the effective potential at $\textstyle\psi=\pm\sqrt{\frac{n}{\xi\left(4-n\right)}}$, so for $\xi>0$ the model is stable for bare potentials which are of quartic order or higher.

\subsection{Bouncing scalar model}
\label{sec:cosmo_BS}

\begin{figure}[t]
	\begin{center}
	{\subfigure[Scale factor]{
		\label{fig:scalar-bounce_scale-factor}
		\includegraphics[height=3.2cm]{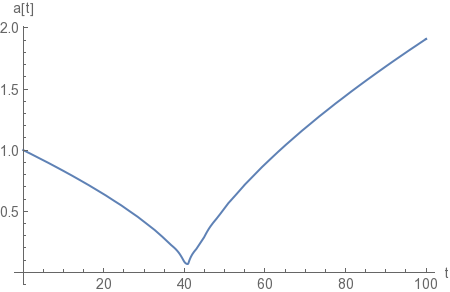}}}
	{\subfigure[Scalar]{
	    \label{fig:scalar-bounce_scalar-field}
		\includegraphics[height=3.2cm]{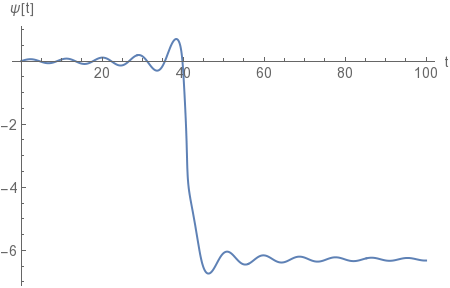}}}
	{\subfigure[Scalar coupling (zoomed)]{
	    \label{fig:scalar-bounce_coupling}
		\includegraphics[height=3.2cm]{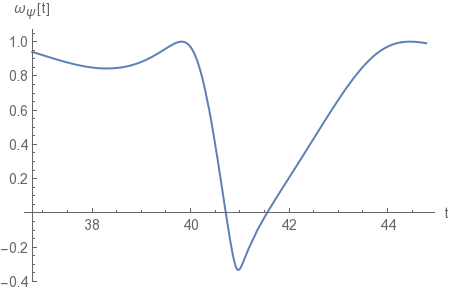}}}
	\end{center}
	\caption{Cosmological bounce generated by non-minimally coupled scalar field with $n=2$ and $\textstyle{}U=\sin^2\left(\psi/2\right)$. Initial conditions, $\psi=0$, $\dot{\psi}=1/25$, $H<0$}
    	\label{fig:scalar-bounce}
\end{figure}

As we said in \subsecref{sec:cosmo_bounce}, we have taken the bounce conditions and constructed a model which bounces purely from the non-minimal coupling.
This model consists of a non-minimally coupled scalar with periodic symmetry.  Our couplings are given by $\omega_R=\cos{\psi}$ and $\textstyle{}\omega_\psi=\frac{1+n\cos{\psi}}{1+n}$, and for simplicity we ignore deformations and the minimally coupled scalar field.  The bouncing scalar model Lagrangian in covariant and effective forms are given by,
\begin{subequations}
\begin{gather}
    L_\mathrm{BS,cov} = N \sqrt{q} \left( \frac{\cos{\psi}}{2} \Rfour - \frac{1+n\cos{\psi}}{2\left(1+n\right)} \partial_\mu \psi \partial^\mu \psi - U \right),
        \\
\begin{aligned}
    L_\mathrm{BS} & = \frac{N \sqrt{q}}{2} \bigg( \cos{\psi} \left( R - \mathcal{K} \right) + \sin{\psi} \left( \nu v + 2 \Delta \psi \right) + \left( \frac{1 + n \cos{\psi}}{1+n} \right) \nu^2 
        \\
    & \quad
    + \left( \frac{ \left(2+n\right) \cos{\psi} - 1}{1+n} \right) \partial_a \psi \partial^a \psi - 2 U \bigg).
\end{aligned}%
\end{gather}%
\end{subequations}
As confirmed by numerically evolving the equations of motion, we know from the bouncing conditions \eqref{eq:cosmo_bounce_conditions} that this model will bounce when $n>1$ because then there is a value of $\psi$ for which $\omega_\psi<0$.
As we show in \figref{fig:scalar-bounce}, the collapsing universe excites the scalar field so much that it `tunnels' through to another minima of the potential.  The bounce happens when the field becomes momentarily ghost-like, when $\omega_\psi<0$.

We can construct other models which produce a bounce purely through non-minimal coupling by having any $U\left(\psi\right)$ with multiple minima and couplings of the approximate form $\omega\sim1-U$.  However, to ensure the scalar does not attempt to tunnel through the potential to infinity and thereby not prevent collapse, the coupling functions must become negative only for values of $\psi$ between stable minima.  For example, for the $\mathbb{Z}_2$ potential
$
    U \left( \psi \right) = \lambda \left( \psi^2 - 1 \right)^2
$,
couplings which are guaranteed to produce a bounce are
$
\omega_R \left( \psi \right) = \omega_\psi \left( \psi \right) = 1 - e^{-\psi^2} \, U \left( \psi \right)
$
when $\lambda > 1$.


\section{Summary}
\label{sec:summary}

In this paper we have presented our calculation of the most general action for a second-order non-minimally coupled scalar-tensor model which satisfies deformed general covariance.  The specific deformation being one motivated by loop quantum cosmology and manifesting as a phase-space function which modifies the canonical constraint algebra.  We presented a similar calculation which involves multiple scalar fields.  We showed how the magnitude of the deformation can be removed by a transformation of the lapse function, but the sign of the deformation and the associated effective signature change cannot be removed.

We explored the background dynamics of the action, in particular showing the conditions required for either a big bounce or a period of slow-roll inflation.  By specifying the free functions we showed how to regain well-known models from our general action.  In particular we discussed the geometric scalar model, which is a parameterisation of $\textstyle{}F\left({}^{(4)}\!R\right)$ gravity and related to the Brans-Dicke model; and we discussed the non-minimally enhanced scalar model of a conventional scalar field with quadratic non-minimal coupling to the curvature.

We presented a model which produces a cosmological bounce purely through non-minimal coupling of a periodic scalar field to gravity.  We also provided the general method of producing similar models without a periodic symmetry.

The investigation into cosmology did not consider in detail the effect that the deformation has on the dynamics.  However, we did show that a big bounce which is purely due to a zeroth order deformation necessarily involves effective signature change.  Further investigation has been left for future work wherein we also include higher order curvature terms in the deformation function.  That is, terms which cannot be factored out by introducing a scalar field such as with the geometric scalar model.


\section*{Acknowledgements}
\label{sec:ackowledgements}

We would like to thank Martin Bojowald for invaluable help during this study.
MS is supported in part by the Science and Technology Facility Council (STFC), UK under the research grant ST/P000258/1.
The work of RC is supported by an STFC studentship.

\appendix


\section{Time asymmetry}
\label{sec:time-asym}

Let us consider time asymmetric terms.  The general Hamiltonian constraint ansatz which includes all terms for a metric field and a single scalar field up to second order in derivatives which are spatially covariant is given by,
\begin{equation}
\begin{split}
    C & = C_\0 + C_{(R)} R + C_{(\psi^{\prime2})} \partial_a \psi \partial^a \psi + C_{(\psi'')} \Delta \psi + C^{(p)} p + C^{(\Pi)} \Pi
    + C^{(p^2)}_{abcd} p^{ab} p^{cd} + C^{(p\Pi)} p \Pi + C^{(\Pi^2)} \Pi^2.
\end{split}
    \label{eq:time-asym_ansatz}
\end{equation}
This does not include any terms such as $\Pi\partial_a\psi$ because they would violate spatial parity and not be spatially covariant.

In the distribution equation \eqref{eq:dist-eqn}, the only contributions coming from the time-asymmetric terms which do not vanish under the $\left(x\leftrightarrow{}y\right)$ symmetry are,
\begin{equation}
    0 = \funcdif{C_0(x)}{q_{ab}(y)} \left( C^{(p)} q^{ab} \right)_y + \funcdif{C_0(x)}{\psi(y)} C^{(\Pi)}(y) - \left( x \leftrightarrow y \right),
        \label{eq:time-asym_dist-eqn}
\end{equation}
which are independent of the terms we consider in \secref{sec:action}.
First order terms in the deformation function appear on their own in the distribution equation if spatial parity is not also violated, and therefore are constrained to vanish so we still only need to consider a zeroth order deformation.

Expanding the functional derivatives in \eqref{eq:time-asym_dist-eqn} into components,
\begin{equation}
\begin{split}
    0 & = \partial_a \delta \left( x, y \right) \left\{ \left. \partdif{C_0}{q_{bc,a}} \right|_x \left( C^{(p)} q^{bc} \right)_y + \left. \partdif{C_0}{\psi_{,a}} \right|_x C^{(\Pi)} (y) \right\}
        \\
    & \quad
    + \partial_{ab} \delta \left( x, y \right) \left\{ \left. \partdif{C_0}{q_{cd,ab}} \right|_x \left( C^{(p)} q^{cd} \right)_y + \left. \partdif{C_0}{\psi_{,ab}} \right|_x C^{(\Pi)} (y) \right\} - \left( x \leftrightarrow y \right),
\end{split}
        \label{eq:time-asym_dist-eqn_2}
\end{equation}
solving using the method shown in \secref{sec:dist-eqn_method-2},
\begin{equation}
    0 = \partdif{C_0}{q_{bc,a}} C^{(p)} q^{bc} + \partdif{C_0}{\psi_{,a}} C^{(\Pi)} + \left( \partial_b C^{(\Pi)} - C^{(\Pi)} \partial_b \right) \partdif{C_0}{\psi_{,ab}} 
    + \left( \partial_b q_{cd} C^{(p)} + q_{cd} \partial_b C^{(p)} - q_{cd} C^{(p)} \partial_b \right) \partdif{C_0}{q_{cd,ab}},
        \label{eq:time-asym_dist-eqn_soln}
\end{equation}
and then substituting in \eqref{eq:C0_derivatives} gives us linearly independent terms,
\begin{equation}
\begin{split}
    0 & = \partial^a \psi \left\{ \left( 2 C_{(\psi^{\prime2})} - \partial_\psi C_{(\psi'')} + C_{(\psi'')} \partial_\psi \right) C^{(\Pi)} + \left( \half C_{(\psi'')} + 2 \partial_\psi C_{(R)} - 2 C_{(R)} \partial_\psi \right) C^{(p)} \right\}
        \\
    & + q^{bc} \partial^a q_{bc} \left\{ \left( \half C_{(\psi'')} - \partial_q C_{(\psi'')} + C_{(\psi'')} \partial_q \right) C^{(\Pi)} + \left( - C_{(R)} + 2 \partial_q C_{(R)} - 2 C_{(R)} \partial_q \right) C^{(p)} \right\}.
\end{split}
\end{equation}
Since the first order terms in the constraint do not affect our calculation of the zeroth and second order terms in \secref{sec:action}, we can substitute in those results here.  Therefore $q^{bc}\partial^aq_{bc}$ can be found to give us the condition,
\begin{equation}
    0 = \left( \partial_q - \frac{\partial_q\beta}{2\beta} \right) \left( C^{(p)} + \frac{\omega_R'}{\omega_R} C^{(\Pi)} \right),
\end{equation}
which is solved by
\begin{equation}
    C^{(p)} = \sigma \sqrt{\left|\beta\right|} - \frac{\omega_R'}{\omega_R} C^{(\Pi)},
\end{equation}
where $\sigma\left(\psi\right)$ is a free function.  Substitute this into the coefficient of $\partial^a\psi$, we find
\begin{equation}
    C^{(\Pi)} = \sqrt{\left|\beta\right|} \left( \frac{ \half \sigma \omega_R' - \sigma' \omega_R }{\omega_\psi + \frac{3\omega_R^{\prime2}}{2\omega_R}} \right).
\end{equation}
If we Legendre transform the constraint in order to find the Lagrangian, we find the additional terms
\begin{equation}
    L \supset \sgn{\beta} N \sqrt{q} \left\{ \half \sigma \omega_R v + \left( \sigma' \omega_R + \sigma \omega_R' \right) \nu \right\}
\end{equation}
which can be combined into $\sgn{\beta}N\mathcal{L}_n\left(\sqrt{q}\sigma\omega_R\right)$ which is a total derivative up to the sign and therefore is a surface term. Hence the time-asymmetric terms do not contribute to the dynamics.


\section{Glossary}
\label{sec:glossary}

\subsection{Definitions}
\label{sec:glossary_definitions}

There are several metric combinations and contractions which are very useful for our calculations,
\begin{equation}
    Q_{abcd} : = q_{a(c} q_{d)b},
        \quad
    Q^{abcd} : = q^{a(c} q^{d)b},
        \quad
    \delta^{ab}_{cd} : = \delta^{a}_{(c} \delta^{b}_{d)},
        \quad
    \partial^{a} : = q^{ab} \, \partial_b,
        \quad
    \Gamma^a : = q^{bc} \, \Gamma^a_{bc}.
        \label{eq:metric_combinations}
\end{equation}
The poisson brackets, for a model with conjugate pairs of variables for a metric field $\textstyle{}\left(q_{ab},p^{ab}\right)$ and multiple scalar fields $\textstyle{}\left(\psi_i,\Pi_i\right)$ are given by,
\begin{equation}
	\left\{ F , G \right\} =
	\int \mathrm{d}^3 x \left( 
	\funcdif{F}{q_{ab}(x)} \funcdif{G}{p^{ab}(x)} 
	+ \sum_i \funcdif{F}{\psi_i(x)} \funcdif{G}{\Pi_i(x)} 
	\right) - \left( F \leftrightarrow G \right).
	    \label{eq:poisson}
\end{equation}
It is useful to define a variable for the Lagrangian density, Hamiltonian constraint and deformation function when time derivatives or momenta are set to zero,
\begin{equation}
	L_0 := \left. L \right|_{ \mathcal{L}_n q_{ab} = \mathcal{L}_n \psi_i = 0 },
	    \quad
	C_0 := \left. C \right|_{ p^{ab} = \Pi_i = 0 },
	    \quad
	\beta_0 := \left. \beta \right|_{ p^{ab} = \Pi_i = 0 }.
        \label{eq:zero_subscript}
\end{equation}
The metric determinant and the logarithmic derivative with respect to the metric determinant appears frequently in our study, so we define for convenience,
\begin{equation}
    q := \det{q_{ab}},
        \quad
    \partial_q := \partdif{}{\,\ln{q}} = q \partdif{}{q}.
        \label{eq:partial_q}
\end{equation}

\subsection{Solving a distribution equation}
\label{sec:dist-eqn_soln}

We present two methods of solving distribution equations which are very useful for the main parts of this paper.

\subsubsection{Method 1}
\label{sec:dist-eqn_method-1}

Look at a model distribution equation with three independent variables to help us calculate the solution,
\begin{equation}
    0 = S \left( x, y \right) \delta \left( z, y \right) + T^a \left( x, y \right) \partial_{a(y)} \delta \left( z, y \right) - \left( x \leftrightarrow y \right),
        \label{eq:dist-eqn_model-1}
\end{equation}
following the procedure in \cite{kuchar_geometrodynamics_1974}, we move the partial derivative and discard the total derivative term,
\begin{equation}
    0 = \left( S \left( x, y \right) - \partial_{a(y)} T^a \left( x, y \right) \right) \delta \left( z, y \right) - \left( x \leftrightarrow y \right),
        \label{eq:dist-eqn_model-1a}
\end{equation}
which we can rewrite as,
\begin{equation}
    0 = U \left( x, y \right) \delta \left( z, y \right) - U \left( y, x \right) \delta \left( z, x \right),
        \label{eq:dist-eqn_model-1b}
\end{equation}
and integrating over $y$,
\begin{equation}
\begin{split}
    0 & = U \left ( x, z \right) - \delta \left( z, x \right) \int \mathrm{d}^3 y \; U \left( y, x \right),
        \\
    & = U \left( x, z \right) - \delta \left( z, x \right) U \left( x \right),
    \quad
    U \left( x \right) = \int \mathrm{d}^3 y \; U \left( y, x \right)
\end{split}
        \label{eq:dist-eqn_model-1c}
\end{equation}
and if we substitute in for $U\left(x,z\right)$, we find our solution to \eqref{eq:dist-eqn_model-1} is given by,
\begin{equation}
    0 = S \left( x, z \right) - \partial_{a(z)} T^a \left( x, z \right) - U \left( x \right) \delta \left( z, x \right),
        \label{eq:dist-eqn_model-1_soln}
\end{equation}
where $U(x)$ is a function which we can calculate, but in practice will not needed when we use this solution.
From here, we multiply by a test function which is a function of $y$, and then integrate over $y$.  The specifics matter for this final step, so we do not write it explicitly here.
This method can easily be generalised to include higher spatial derivatives if needed.

\subsubsection{Method 2}
\label{sec:dist-eqn_method-2}

Look at a model distribution equation with two independent variables to help us calculate the solutions,
\begin{equation}
    0 = \partial_a \delta \left( x, y \right) S^a (x) T (y) + \partial_{ab} \delta \left( x, y \right) U^{ab} (x) V (y) - \left( x \leftrightarrow y \right),
        \label{eq:dist-eqn_model-2}
\end{equation}
multiply by test functions $t(x)\tau(y)$ and integrate over $x$ and $y$,
\begin{equation}
\begin{split}
    0 & = 
    \int \mathrm{d}^3 y \left( \tau T \right)_y
    \int \mathrm{d}^3 x \left( t S^a \partial_a \right)_x
    \delta \left( x, y \right)
    - \int \mathrm{d}^3 x \left( t T \right)_x
    \int \mathrm{d}^3 y \left( \tau S^a \partial_a \right)_y \delta \left( y, x \right)
        \\ & \quad
    \int \mathrm{d}^3 y \left( \tau V \right)_y
    \int \mathrm{d}^3 x \left( t U^{ab} \partial_{ab} \right)_x
    \delta \left( x, y \right)
    - \int \mathrm{d}^3 x \left( t V \right)_x
    \int \mathrm{d}^3 y \left( \tau U^{ab} \partial_{ab} \right)_y \delta \left( y, x \right),
\end{split}
    \label{eq:dist-eqn_model-2a}
\end{equation}
and integrate by parts to find separable terms,
\begin{equation}
    0 = \int \mathrm{d}^3x \left\{ \left( \tau \partial_a t - t \partial_a \tau \right) \left[ S^a T + \left( \partial_b V - V \partial_b \right) U^{ab} \right]
    + 2 \partial_{[a} \tau \, \partial_{b]} t \, U^{ab} V \right\},
        \label{eq:dist-eqn_model-2b}
\end{equation}
and therefore the solution to \eqref{eq:dist-eqn_model-2} is given by,
\begin{equation}
    0 = S^a T + \left( \partial_b V - V \partial_b \right) U^{ab},
        \quad
    0 = U^{[ab]} V,
        \label{eq:dist-eqn_model-2_soln}
\end{equation}
this cannot easily be generalised to include higher spatial derivatives.  In that circumstance, it would be easier to use Method 1 in \secref{sec:dist-eqn_method-1}.

\subsection{Variation of the Riemann tensor}
\label{sec:variation_of_riemann}

The variation of the Riemann tensor is given by the Palatini equation,
\begin{equation}
	\delta R^a_{\;\,bcd} = \nabla_c \delta \Gamma^a_{db} - \nabla_d \delta \Gamma^a_{cb},
	    \label{eq:var_riemann_1}
\end{equation}
where the variation of the Levi-Civita connection is
\begin{equation}
	\delta \Gamma^a_{bc} = 
	\half q^{ad} \left(
		\nabla_b \delta q_{dc} + \nabla_c \delta q_{bd} - \nabla_d \delta q_{bc}
	\right)
	    \label{eq:var_connection}
\end{equation}
from which we can deduce,
\begin{equation}
	\delta R^a_{\;\,bcd} = 
	\Theta^{a \;\;\;\;\; ef}_{\;\,bcd} \delta q_{ef}
	+ \Phi^{a \;\;\;\;\; efgh}_{\;\,bcd} \nabla_{ef} \delta q_{gh}
	    \label{eq:var_riemann_2}
\end{equation}
where we have defined the useful tensors,
\begin{subequations}
\begin{align}
	\Theta^{a \;\;\;\;\; ef}_{\;\;bcd} & =
	\frac{-1}{2} \left( q^{a(e} R^{f)}_{\;\;\;\;bcd} + \delta^{(e}_b R^{f)a}_{\;\;\;\;\;\;\;cd} \right),
	    \label{eq:var_coeff_Theta} \\
	\Phi^{a \;\;\;\;\; efgh}_{\;\;bcd} & = \half \left( q^{a(e} \delta^{f)}_d \delta^{gh}_{bc} + q^{a(g} \delta^{h)}_d \delta^{ef}_{bc} - q^{a(e} \delta^{f)}_c \delta^{gh}_{bd} - q^{a(g} \delta^{h)}_c \delta^{ef}_{bd} \right),
	    \label{eq:var_coeff_Phi}
\end{align}
    \label{eq:var_coeff}%
\end{subequations}
\!\!but contracted versions of these are more useful,
\begin{subequations}
\begin{align}
	\Theta^{cd}_{ab}
	& := \delta^{ef}_{ab} \Theta^{g \;\;\;\;\; cd}_{\;\;egf} = \half \left( Q^{cdef} R_{e(ab)f} + \delta^{(c}_{(a} R^{d)}_{b)}	\right),
	\quad q_{cd} \Theta^{cd}_{ab} = q^{ab} \Theta^{cd}_{ab} = 0,
	    \label{eq:var_coeff_contract_Theta} \\
	\Phi_{ab}^{cdef}
	& := \delta^{gh}_{ab} \Phi^{i \;\;\;\;\; cdef}_{\;\;gih} = \half \left( q^{c(e} \delta^{f)d}_{ab} + q^{d(e} \delta^{f)c}_{ab} - q^{cd} \delta^{ef}_{ab} - q^{ef} \delta^{cd}_{ab} \right),
	    \label{eq:var_coeff_contract_Phi} \\
	\Phi^{abcd}
	& := q^{ef} \Phi^{abcd}_{ef} = Q^{abcd} - q^{ab} q^{cd}.
	    \label{eq:var_coeff_dblcontract_Phi}
\end{align}%
    \label{eq:var_coeff_contract}%
\end{subequations}
To decompose the Riemann tensor in terms of partial derivatives, use this formula for decomposing the second covariant derivative of the variation of the metric,
\begin{equation}
\begin{split}
    \nabla_d \nabla_c \delta q_{ab}
    & = \partial_d \partial_c \delta q_{ab} + \partial_g \delta q_{ef} \left( - \Gamma^g_{dc} \delta^{ef}_{ab} - 4 \delta^{(e}_{(a} \Gamma^{f)}_{b)(c} \delta^g_{d)} \right)
        \\
    & \quad
    + \delta q_{ef} \left( - 2 \partial_d \Gamma^{(e}_{c(a} \delta^{f)}_{b)} + 2 \Gamma^g_{dc} \Gamma^{(e}_{g(a} \delta^{f)}_{b)} + 2 \Gamma^g_{d(a} \delta^{(e}_{b)} \Gamma^{f)}_{cg} + 2 \Gamma^{(e}_{d(a} \Gamma^{f)}_{b)c} \right).
\end{split}
    \label{eq:var_metric}
\end{equation}


\bibliographystyle{./dgrstmbib}
\bibliography{./dgrstm}

\end{document}